\documentclass[a4paper]{jpconf}

\usepackage{lmodern}
\usepackage{amsmath}
\usepackage{amssymb}
\usepackage{graphicx}
\usepackage{nomencl}
\usepackage{cite}
\usepackage{float}
\usepackage[pdftex]{hyperref}

\PassOptionsToPackage{hyphens}{url}\usepackage{hyperref}

\begin{document}

\title{Circularly Symmetric Light Waves: An Overview}
\author{Andrea Cagliero$^1$}
\address{$^1$ Microwaves Department of IMT Atlantique, Institut Mines-T\'{e}l\'{e}com; Lab-STICC (CNRS), Laboratory for Science and Technologies of Information, Communication and Knowledge, Brest, F-29238, France}
\ead{andrea.cagliero@edu.unito.it}

\address{\textbf{\\The present manuscript has been accepted for publication in ``Journal of Optics'', DOI: 10.1088/2040-8986/aad113. It is available for reuse under a CC BY-NC-ND 3.0 licence after the 12 month embargo period provided that all the terms of the licence are adhered to.}}

\begin{abstract}
Orbital Angular Momentum (OAM) waves were first recognized as those specific vortex solutions
of the paraxial Helmholtz equation for which the orbital contribution
to the total angular momentum of the beam yields an integer multiple
of $\hbar$ along the propagation direction. However, this class of
solutions can be generalized to include more sophisticated vector
vortex waves with coupled polarization and spatial complexity, that
are eigenfunctions of the third component of the angular momentum
operator. In this work, a rigorous framework is proposed for the analysis of
all the possible families of vortex solutions to the homogeneous Helmholtz
equation. Both the scalar and vector cases are studied in depth, making
use of an operator approach which emphasizes their intimate connection
with the two-dimensional rotation group. Furthermore, a special focus
is given to the characterization of the propagation properties of
the most popular families of paraxial OAM beams.
\end{abstract}

\section{Introduction}

Electromagnetic waves carry energy and both linear and angular momenta.
Whereas a contribution to the total angular momentum is realized by
the spin of the photon, the fundamental physical quantity associated
with polarization, an orbital angular momentum (OAM\nomenclature{OAM}{orbital angular momentum})
component can also be present. Such contribution is often said to
be ``quasi-intrinsic'' \cite{Zambrini2006}, since it is independent
of the axis of calculation for any beam endowed with a helical wavefront
apertured symmetrically about the beam axis \cite{ONeil2002}. The
OAM content of helical beams in the paraxial regime was first discovered
by Les Allen et al. in 1992 \cite{Allen1992}, although these waves
have been known for much longer in terms of optical vortices and are
now often referred to as vortex waves. OAM beams are characterized 
by an azimuthal phase dependence of the form $\exp\left(im\varphi\right)$, 
where the integer $m$ represents the topological charge of the vortex 
(or, more simply, the number of intertwined helices the wavefront is made of) 
and is related to the OAM carried by the beam along the propagation axis. For $m\neq0$,
due to the presence of the on-axis phase singularity, scalar OAM waves
present a doughnut-shaped intensity profile with a central null. 

It would seem natural to think that, being the OAM a global property
of the beam in contrast to the spin angular momentum (SAM\nomenclature{SAM}{spin angular momentum}),
the orbital and spin contributions to the total angular momentum of
an electromagnetic wave are two distinct quantities. However, it has
been argued that, in view of the absence of any rest frame for the
photon, a gauge-invariant decomposition of the total electromagnetic
angular momentum in orbital and spin parts is unfeasible (see for
example \cite{Jauch1976,Simmons1970}) or achievable at most within
the paraxial limit \cite{Barnett1994}, the polarization and spatial
degrees of freedom being in general coupled. Despite these arguments,
Steven J. van Enk and Gerald Nienhuis have claimed that it is indeed
possible to extract gauge-independent expressions for the spin and
orbital parts of the angular momentum, but neither of these contributions
alone represents a true angular momentum, since the transversality
of the radiation fields affects the commutation relations for the
associated quantum operators \cite{VanEnk1994}. Such interpretation
proves to be the currently accepted one \cite{Barnett2016}.

The first experimental proof of an angular momentum transfer between
light and matter dates back to 1936, when Richard A. Beth verified
the interaction between a circularly polarized beam and a quarter-wave
plate \cite{Beth1936}: owing to the photon spin, the torque exerted
by the polarized radiation on the plate put it into rotation. Since
then, a great number of similar experiments have been performed, also
involving the use of OAM beams. The angular momentum of light can
be transferred to suitable trapped material particles causing them
to rotate \cite{Friese1996,Simpson1997}, with relevant applications
in both micromanipulation and the design and operation of micromachines
\cite{Galajda2001,Grier2003,Padgett2011,Paterson2001}. Vortex waves
are also used as a tool for advanced cell manipulation \cite{Jeffries2007}
and to transport aqueous droplets in solution \cite{Lorenz2006}.
Moreover, direct transfer of OAM from light to atoms in ultracold
gas clouds has been reported \cite{Andersen2006,Tabosa1999,Wright2008}.

The interest in OAM waves extends far beyond the light-matter angular
momentum exchange and several attractive applications can be found
in optics, quantum physics and astronomy. For instance, light vortices
are employed to probe different physical and biological properties
of matter in optical imaging \cite{Bernet2006,Furhapter2005,Wang2015},
but also for encoding quantum information via the corresponding photon
states in high-dimensional Hilbert spaces \cite{Dada2011,Fickler2012,Mair2001,MolinaTerriza2007,Vaziri2002}.
OAM beams are involved in second harmonic generation processes \cite{Bovino2011,Dholakia1996,Fang2015}
as well as in many other nonlinear optics phenomena \cite{Bigelow2004,Ferrando2004,Firth1997,Minardi2001},
being intended in terms of vortex solitons \cite{Kruglov1985,Law1993,Swartzlander1992}.
Further possibilities lie in the use of vortex waves to enhance the
resolution in quantum imaging \cite{Jack2009} and for overcoming
the Rayleigh limit of the telescopes \cite{Tamburini2006}, whereas
several other interesting studies are reported by the astrophysical
community \cite{Berkhout2008,Elias2008,Foo2005,Harwit2003,Tamburini2011}.
Many of the so far described applications together with other examples
and a more complete list of references can be found in \cite{Allen2003,Andrews2008,Andrews2013,RubinszteinDunlop2016,Torres2011,Yao2011}. 

Currently, a large part of the research on OAM beams is also devoted
to the field of telecommunications. From a strictly mathematical point
of view, vortex waves represent classes of solutions to the Helmholtz
equation; since each class comprises a set of orthogonal solutions,
in 2004 Graham Gibson et al. proposed the idea of using such waves
to convey independent information channels on a single frequency in
free-space optical communication links \cite{Gibson2004}. The authors
also suggested how the intrinsic sensitivity of the OAM orthogonality
to angular restrictions and lateral offsets \cite{FrankeArnold2004}
could prevent eavesdropping. 

In these last few years, a strong interest in the possibility to increase
the communication efficiency by vortex waves is growing \cite{Anguita2008,Bozinovic2013,Djordjevic2011,Qu2016,Wang2012,Willner2015,Zhu2016}
and has also been extended to the radio frequency domain \cite{Gaffoglio2016,Mohammadi2010,Tamburini2012,Thide2007,Yan2014}.
In this scenario, however, the degree of innovation of OAM multiplexing
with respect to other existing techniques like those based on antenna
diversity and, more generally, on the use of orthogonal wavefields
has been questioned \cite{Andersson2015,Berglind2014,Chen2016,Edfors2012,Gaffoglio2017,Tamagnone2012,Tamagnone2013}. 

Motivated by the astonishing amount of OAM-related publications, 
a theoretical study is proposed in this paper on the mathematical 
features of free-space vortex waves. While presenting new results 
and general insights that can be of interest to the optics community, 
a large part of the literature on this fascinating topic is carefully reviewed. 
In Section \ref{sec:Scalar}, a detailed overview on the scalar OAM waves is outlined, 
starting from some basic results of group theory. In Section \ref{sec:Paraxial},
the fundamental solutions of the paraxial Helmholtz equation are derived
step by step through an ansatz approach and their propagation properties
are systematically analyzed. In Section \ref{sec:Vector}, the vector
case is studied in depth from a new perspective and some well-known examples
are revisited accordingly. The last section is devoted to the conclusions. 
Further calculations are reported in appendix as a support to the main text. 

\section{Vortex solutions of the scalar wave equation\label{sec:Scalar}}

As it is known in group theory, separable coordinate systems for second-order
linear partial differential equations can be characterized in terms
of sets of operators in the algebra relative to the underlying continuous
symmetry group \cite{Miller1977}. In this framework, the separated
solutions of the considered equations are common eigenfunctions of
the symmetry operators and the expansion of one set of separable solutions
in terms of another leads back to a problem in the representation
theory of the Lie algebra. The Helmholtz wave equation, one of the
most studied examples, is proven to be separable in eleven three-dimensional
coordinate systems \cite{Morse1953}: Cartesian, circular cylindrical,
elliptic cylindrical, parabolic cylindrical, spherical, prolate spheroidal,
oblate spheroidal, parabolic, ellipsoidal, paraboloidal and conical.
In optics, solutions of the Helmholtz equation which are eigenfunctions
of the Lie algebra generator of the translations along the $z$-coordinate
are usually considered, since only for them the equation can be separated
into transverse and longitudinal parts; actually, this condition is
satisfied in the first four orthogonal coordinate systems.

A useful approach for building a formal description of OAM waves is
to analyze their definition from a very general point of view, proceeding
along the lines of the methods and techniques developed in group theory.
The fundamental feature which has been referred to in defining vortex
waves from the very beginning is indeed represented by the aforementioned
screw phase dislocation $\exp\left(im\varphi\right)$, whose presence
is soon recognized as the footprint of the underlying circular symmetry.
In particular, given a fixed axis along the unit vector
$\mathbf{u}_{z}$, rotations about this axis form a one-parameter
subgroup of SO(3)\nomenclature{SO(3)}{rotation group in the three-dimensional Euclidean space}
which is isomorphic to the group of rotations in the plane perpendicular
to $\mathbf{u}_{z}$, that is SO(2)\nomenclature{SO(2)}{rotation group in the two-dimensional Euclidean space}
\cite{Tung1985}. Through the azimuthal integer index $m$, the function
$\varPhi_{m}\left(\varphi\right)=\exp\left(im\varphi\right)$ is known
to span the set of single-valued irreducible representations of SO(2),
for which the following orthogonality and completeness relations hold:
\begin{equation}
\frac{1}{2\pi}\int_{0}^{2\pi}\varPhi_{j}^{*}\left(\varphi\right)\varPhi_{m}\left(\varphi\right)\,d\varphi=\delta_{jm};\quad\sum_{m}\varPhi_{m}\left(\varphi\right)\varPhi_{m}^{*}\left(\varphi'\right)=\delta\left(\varphi-\varphi'\right).\label{eq:orthogonalityCompleteness}
\end{equation}
Associated with the subgroup algebra there is a generator, $\hat{J}_{z}$,
and all elements of the given subgroup can be written symbolically
as $\exp\left(\varphi\hat{J}_{z}\right)$, where $\varphi$ parameterizes
the rotation. 

A basis for the Lie algebra of SO(3) in standard matrix
representation is provided by the three operators:
\begin{equation}
J_{1}=\left[\begin{array}{ccc}
0 & 0 & 0\\
0 & 0 & -1\\
0 & 1 & 0
\end{array}\right];\quad J_{2}=\left[\begin{array}{ccc}
0 & 0 & 1\\
0 & 0 & 0\\
-1 & 0 & 0
\end{array}\right];\quad J_{3}=\left[\begin{array}{ccc}
0 & -1 & 0\\
1 & 0 & 0\\
0 & 0 & 0
\end{array}\right],\label{eq:Jk}
\end{equation}
which obey the commutation relations $\left[J_{s},\,J_{v}\right]={\ensuremath{\varepsilon_{sv}}}^{k}J_{k}$,
where $\ensuremath{\varepsilon_{svk}}$ is the Levi-Civita pseudo-tensor
and the Einstein summation convention has been considered; in (\ref{eq:Jk}),
the subscripts $1$, $2$ and $3$ denote Cartesian components $x$,
$y$ and $z$, respectively. Although the origin of these commutation
relations is really geometric in nature, the generators acquire even
more significance in quantum physics, where they correspond to measurable
quantities \cite{Messiah1961}. In this respect, $\mathbf{J}=\left(J_{1},J_{2},J_{3}\right)$
can be seen as the vector angular momentum operator in units of $\hbar$. 

Whereas the matrix representation is widely used to express the SAM
operator, a differential representation is often preferable to (\ref{eq:Jk})
for describing the OAM operator. As suggested in \cite{Nienhuis1993},
analogies between quantum mechanics and scalar paraxial optics can
be built in order to demonstrate heuristically that some cylindrical
laser modes with a $\varPhi_{m}\left(\varphi\right)$ azimuthal phase
dependence carry an OAM of $m\hbar$ per photon along the propagation
axis. 

Let us consider an arbitrary coordinate system $\left\{ x_{1},x_{2},x_{3}\right\} $
and a free-space monochromatic scalar wave of the form $\Psi\left(x_{1},x_{2},x_{3},t\right)=\psi\left(x_{1},x_{2},x_{3}\right)\mathrm{e}^{i\omega t}$,
where $\omega=kc$ represents the angular frequency, $k=2\pi/\lambda$
the wavenumber, $\lambda$ the wavelength and $c$ the speed of light.
Even if the full vector nature of the electromagnetic radiation is
being neglected at present, $\Psi\left(x_{1},x_{2},x_{3},t\right)$
can be understood as the amplitude of a linearly polarized electric
field or vector potential, which satisfies the wave equation:
\begin{equation}
\nabla^{2}\Psi\left(x_{1},x_{2},x_{3},t\right)-\frac{1}{c^{2}}\frac{\partial^{2}}{\partial t^{2}}\Psi\left(x_{1},x_{2},x_{3},t\right)=0,\label{eq:waveEquation}
\end{equation}
as follows from Maxwell theory. From (\ref{eq:waveEquation}), the
scalar Helmholtz equation for the spatial coordinates dependent amplitude
$\psi\left(x_{1},x_{2},x_{3}\right)$ is soon derived:
\begin{equation}
\nabla^{2}\psi\left(x_{1},x_{2},x_{3}\right)+k^{2}\psi\left(x_{1},x_{2},x_{3}\right)=0.\label{eq:HelmholtzEquation}
\end{equation}
We are now interested in solutions $\psi$ to (\ref{eq:HelmholtzEquation})
that are eigenfunctions of the $\hat{J}_{z}$ operator, namely: 
\begin{equation}
i\hat{J}_{z}\psi=-i\frac{\partial}{\partial\varphi}\psi=m\psi\quad\Rightarrow\quad\psi\left(x_{1},\varphi,x_{3}\right)=\varPhi_{m}\left(\varphi\right)u\left(x_{1},x_{3}\right),\label{eq:PhiSeparation}
\end{equation}
where $\left\{ x_{1},\varphi,x_{3}\right\} $ is a suitable coordinate
system and the differential expression of the $\hat{J}_{z}$ operator has
been introduced. It is important to note that the condition of periodicity
of the wavefunction in the azimuthal coordinate imposes integer values
for the $m$ index. In (\ref{eq:PhiSeparation}), $\varPhi_{m}\left(\varphi\right)$
corresponds to the sought vortex term and $u\left(x_{1},x_{3}\right)$
is a function obeying the reduced equation in the remaining two variables:
\begin{equation}
\frac{1}{h_{1}h_{\varphi}h_{3}}\left[\frac{\partial}{\partial x_{1}}\left(\frac{h_{\varphi}h_{3}}{h_{1}}\frac{\partial u}{\partial x_{1}}\right)+\frac{\partial}{\partial x_{3}}\left(\frac{h_{1}h_{\varphi}}{h_{3}}\frac{\partial u}{\partial x_{3}}\right)\right]+\left(k^{2}-\frac{m^{2}}{h_{\varphi}^{2}}\right)u=0,\label{eq:reducedEquation}
\end{equation}
being $h_{1}$, $h_{\varphi}$ and $h_{3}$ three $\varphi$-independent
metric scale factors for the given coordinates \cite{Morse1953}.

It can be shown that (\ref{eq:reducedEquation}) separates in five
coordinate systems: circular cylindrical, spherical, prolate spheroidal,
oblate spheroidal and parabolic \cite{Miller1977}. While leaving
to the appendix a formal proof of the equation separation in the latter
four systems, let us focus in detail on the circular cylindrical case
$\left\{ \rho,\varphi,z\right\} $, for which equation (\ref{eq:reducedEquation})
reads: 
\begin{equation}
\left[\frac{\partial^{2}}{\partial\rho^{2}}+\frac{1}{\rho}\frac{\partial}{\partial\rho}-\frac{m^{2}}{\rho^{2}}+\frac{\partial^{2}}{\partial z^{2}}+k^{2}\right]u\left(\rho,z\right)=0.\label{eq:reducedEquationCylindrical}
\end{equation}
As pointed out above, it is common to look for solutions of the Helmholtz
equation that are eigenfunctions of the $\hat{P}_{z}$ operator, i.e.
the generator of the translations along the $z$-axis:
\begin{equation}
\hat{P}_{z}u=\frac{\partial}{\partial z}u=-ik_{z}u\quad\Rightarrow\quad u\left(\rho,z\right)=B\left(\rho\right)e^{-ik_{z}z},
\end{equation}
where $B\left(\rho\right)$ represents a function to be determined.
The imposed condition implies a further reduction of equation (\ref{eq:reducedEquationCylindrical}):
\begin{equation}
\left[\frac{\partial^{2}}{\partial\rho^{2}}+\frac{1}{\rho}\frac{\partial}{\partial\rho}-\frac{m^{2}}{\rho^{2}}+\left(k^{2}-k_{z}^{2}\right)\right]B\left(\rho\right)=0.\label{eq:BesselEquation}
\end{equation}
After some straightforward manipulations, (\ref{eq:BesselEquation})
is easily recognized as the Bessel equation (see, for instance, \cite{Abramowitz1972})
and thus $B\left(\rho\right)=J_{\left|m\right|}\left(k_{\rho}\rho\right)$,
where $k^{2}=k_{\rho}^{2}+k_{z}^{2}$. 

\begin{figure}
\noindent \begin{centering}
\includegraphics[width=1\textwidth]{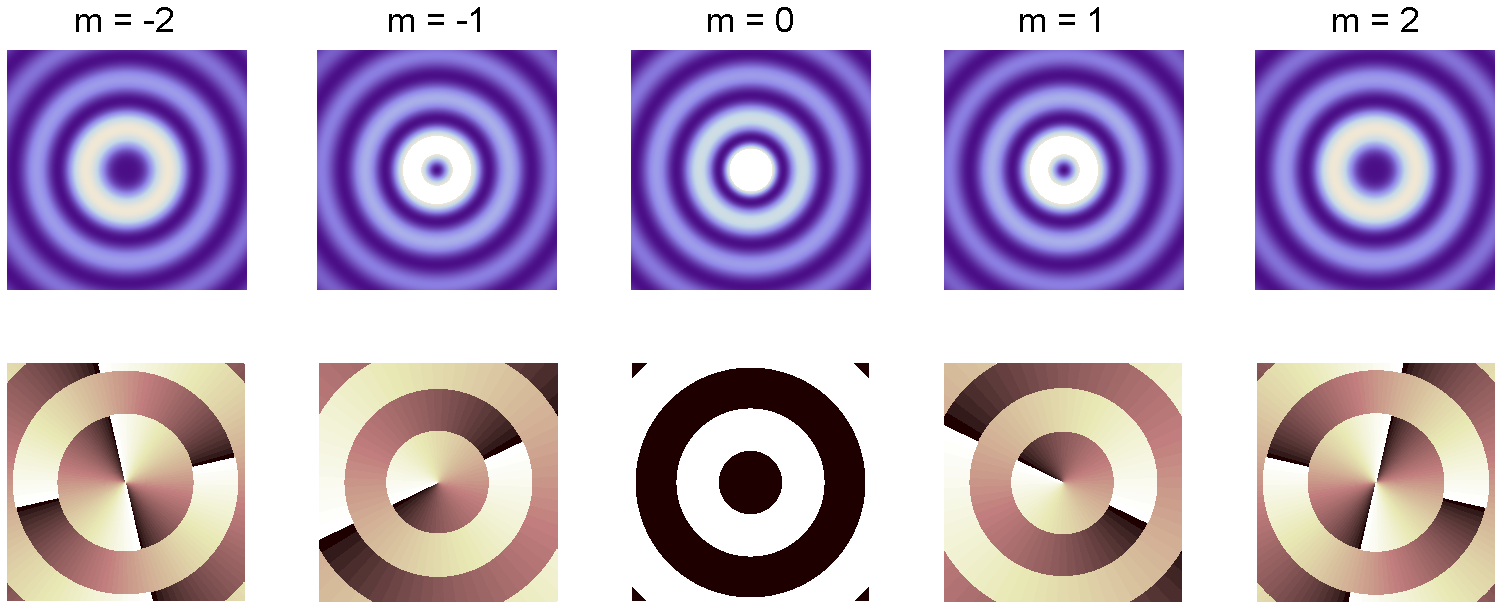}
\par\end{centering}
\caption{BB intensity (upper row) and phase (lower row) transverse profiles ($y$-coordinate versus $x$-coordinate) for
$k_{\rho}=k/\sqrt{2}$.\label{fig:BBs}}
\end{figure}

Scalar modes of the form: 
\begin{equation}
\varPsi_{m}^{B}\left(\rho,\varphi,z,t;k_{\rho}\right)=C_{m}^{B}J_{\left|m\right|}\left(k_{\rho}\rho\right)\varPhi_{m}\left(\varphi\right)\exp\left(-ik_{z}z+i\omega t\right),\label{eq:BesselModes}
\end{equation}
where $C_{m}^{B}$ is a suitable dimensional constant, are known as
\textit{Bessel beams} (BBs\nomenclature{BB}{Bessel beam}) and represent
a complete set of orthogonal solutions to (\ref{eq:waveEquation}). The transverse intensity and phase profiles of a representative BB
are displayed in Figure \ref{fig:BBs} for different values of the
topological charge $m$.
As will be shown further on, BBs also satisfy the paraxial wave equation.
Due to their divergence-free profile, the Bessel modes belong to the
family of ``non-diffracting'' beams \cite{Durnin1987}, which are
not square-integrable and thus carry infinite energy; for this reason,
only truncated forms of such waves can be realized experimentally.
Physical OAM waves that approximate the intensity distribution of
these ideal beams are usually produced via computer generated holograms
\cite{Paterson1996} or axicons \cite{Arlt2000}. 

It is important to emphasize that, despite the leading role played
by the circular cylindrical coordinate system in optics, BBs are not
the only possible scalar OAM solutions and waves with a vortex term
$\varPhi_{m}\left(\varphi\right)$ are also found by separating the
Helmholtz equation in spherical, prolate spheroidal, oblate spheroidal
and parabolic coordinates (see \ref{sec:A1} for further details).
In \cite{ChavezCerda2002}, an estimation of the electromagnetic OAM
has also been performed for Mathieu beams, which can be intended as
a generalization of BBs to elliptic cylindrical coordinates; in this
case, however, the $\varPhi_{m}\left(\varphi\right)$ term cannot
be present due to symmetry considerations and this results in the
appearance of a more complicated vortex structure with fractional
OAM mean content. 

The presence of a non-integer OAM per photon in connection with some
sort of symmetry breaking is a fact, by the way, not new, that leads
to possible extensions of the simple original definition of OAM waves
in several interesting scenarios \cite{Berry2004,Galvez2007,Gotte2007,Gotte2008,Leach2004,Nugrowati2012,ODwyer2010,Oemrawsingh2005}.
As a related topic, it should be noted that the very concept of wave
vorticity is not limited to the isotropic case, which has been taken
as a working hypothesis in defining OAM beams as vortex solutions.
In particular, a vortex is said to be \textit{isotropic} when the
phase increases linearly from $0$ to $2\pi m$ around a circle enclosing
the singularity \cite{Dickey2014,Wolf2009}; in this case, the iso-intensity
contour lines are exact circles and the usual $\varPhi_{m}\left(\varphi\right)$
term is found. However, vortices are in general \textit{anisotropic},
meaning that the intensity contours close to the singularities are
elliptical \cite{ChavezCerda2001,Freund1997,LopezMariscal2006,LopezMariscal2007,MolinaTerriza2001,Schechner1996}.
The anisotropic vortex can be understood as a geometrical deformation
of the isotropic one, with an angular dependence of the form:
\begin{equation}
\widetilde{\varPhi}_{m}\left(\varphi\right)=\sqrt{2}\left\{ \cos\epsilon\cos\left[m\left(\varphi-\varphi_{0}\right)\right]+i\sin\epsilon\sin\left[m\left(\varphi-\varphi_{0}-\sigma\right)\right]\right\} ,
\end{equation}
where $\epsilon$ represents the vortex anisotropy, $\sigma$ is the
vortex skewness and $\varphi_{0}$ the vortex rotation angle. The
possibility of representing anisotropic dislocations through a superposition
of isotropic ones, as reported in \cite{Schechner1996}, provides
an easy derivation of the expectation value and the uncertainty of
the OAM. 

A further aspect to be mentioned lies in the non-existence of propagating
solutions characterized by a term $\varPhi_{\mu}\left(\varphi\right)$
with non-integer $\mu$ index: when programmed to have a non-integer
phase dislocation of this form as a starting condition, electromagnetic
beams evolve with the creation of a group of standard vortices which
break the circular symmetry of the intensity profile. Such mechanism
and all the related topics have been deeply explored and characterized
\cite{Berry2004,Galvez2007,Gotte2008,Leach2004,Wolf2009}. 

The study of the electromagnetic vortices and their dynamics and interactions
belong to the rich field of the singular optics, to which reference
should be made. In spite of all the possible generalizations that
can be introduced in the definition of OAM waves as vortex solutions,
only the original requirement, i.e. the presence of an on-axis isotropic
vortex, will be taken into account, since a complete classification
would fall beyond the scope of the paper.

\section{Paraxial OAM waves\label{sec:Paraxial}}

In the event that the angle between the wavevector $\mathbf{k}$
and the propagation direction ($z$-axis) is small and assuming a
beam amplitude of the form $\psi\left(\rho,\varphi,z\right)=f\left(\rho,\varphi,z\right)\exp\left(-ikz\right)$,
where $f\left(\rho,\varphi,z\right)$ represents a slowly varying
function of the $z$-coordinate, such that: 
\begin{equation}
\left|\frac{\partial^{2}f\left(\rho,\varphi,z\right)}{\partial z^{2}}\right|\ll k\left|\frac{\partial f\left(\rho,\varphi,z\right)}{\partial z}\right|;\quad\left|\frac{\partial^{2}f\left(\rho,\varphi,z\right)}{\partial z^{2}}\right|\ll\left|\nabla_{T}^{2}\:f\left(\rho,\varphi,z\right)\right|,
\end{equation}
being $\nabla_{T}^{2}$ the transverse Laplacian, equation (\ref{eq:HelmholtzEquation})
reduces to its paraxial version \cite{Kogelnik1966}: 
\begin{equation}
\left[\frac{\partial^{2}}{\partial\rho^{2}}+\frac{1}{\rho}\frac{\partial}{\partial\rho}+\frac{1}{\rho^{2}}\frac{\partial^{2}}{\partial\varphi^{2}}-2ik\frac{\partial}{\partial z}\right]f\left(\rho,\varphi,z\right)=0.\label{eq:paraxialHelmholtzEquation}
\end{equation}
The paraxial Helmholtz equation (\ref{eq:paraxialHelmholtzEquation})
is widely used in optics, where the above described requirements are
usually met and it represents a complete enough approximation to handle
transverse variations and diffraction effects of the optical beam
profile \cite{Siegman1986}. The reported equation is of the Schr\"{o}dinger
type and proves to admit solutions with separable variables in seventeen
coordinate systems \cite{Miller1977}. 

A huge amount of publications regarding the analytical derivation
of families of paraxial beams can be found in the literature \cite{Abramochkin2004,Bandres2004b,Bandres2004a,Bandres2007,Bandres2008a,Bandres2008b,Bandres2010,Caron1999,Casperson1997,Casperson1998,Gori1987,GutierrezVega2007a,GutierrezVega2007b,GutierrezVega2005,Karimi2008,Karimi2007,Khonina2011,Kotlyar2006,Kotlyar2008,Kotlyar2007,Li2004,Porras2001,Pratesi1977,Siegman1973,Siviloglou2007a,Siviloglou2007b,Sun2012,Wunsche1989,Zauderer1986}.
The most general set of solutions of (\ref{eq:paraxialHelmholtzEquation})
in circular cylindrical coordinates has been obtained and characterized
recently \cite{Bandres2008a,Bandres2010}; closed form expressions
are provided in terms of the confluent hypergeometric functions \cite{Abramowitz1972}.
Among all special cases of the paraxial vortex waves, some cylindrical
families are of particular relevance and will be analyzed by means
of an intuitive ansatz approach \cite{Karimi2009}. 

First, in order to simplify the notation, it is useful to introduce
a set of dimensionless circular cylindrical coordinates $\left\{ \varrho,\varphi,\zeta\right\} =\left\{ \rho/w_{0},\varphi,z/z_{R}\right\} $,
where $z_{R}=\pi w_{0}^{2}/\lambda$ and $w_{0}$ corresponds to a
characteristic length parameter for the beam. Equation (\ref{eq:paraxialHelmholtzEquation})
then becomes:
\begin{equation}
\left[\frac{\partial^{2}}{\partial\varrho^{2}}+\frac{1}{\varrho}\frac{\partial}{\partial\varrho}+\frac{1}{\varrho^{2}}\frac{\partial^{2}}{\partial\varphi^{2}}-4i\frac{\partial}{\partial\zeta}\right]f\left(\varrho,\varphi,\zeta\right)=0.\label{eq:adimensionalParaxialEquation}
\end{equation}
For cylindrical beams satisfying the requirements of the paraxial
approximation it is useful to appeal to the following ansatz, which
is eigenfunction of the $\hat{J}_{z}$ operator by definition:
\begin{equation}
f\left(\varrho,\varphi,\zeta\right)=\frac{C}{\mu\left(\zeta\right)}u\left[\frac{\varrho}{\mu\left(\zeta\right)},\zeta\right]\exp\left[i\phi\left(\zeta\right)\frac{\varrho^{2}}{\mu^{2}\left(\zeta\right)}+im\varphi\right],\label{eq:Ansatz}
\end{equation}
where $\mu\left(\zeta\right)$ and $\phi\left(\zeta\right)$ are two
dimensionless functions which account for the diffraction effects,
$m\in\mathbb{Z}$ represents the topological charge of the on-axis
vortex and $C$ is a constant. By inserting expression
(\ref{eq:Ansatz}) into (\ref{eq:adimensionalParaxialEquation}),
a partial differential equation for the function $u\left(r,\zeta\right)$
is found which can be put in the form of a $\zeta$-dependent Schr\"{o}dinger
equation:
\begin{equation}
i\frac{\partial u\left(r,\zeta\right)}{\partial\zeta}=\hat{H}\left(\zeta\right)u\left(r,\zeta\right)\label{eq:SchrodingerEquation}
\end{equation}
with Hamiltonian given by:
\begin{eqnarray}
\hat{H}\left(\zeta\right) & = & \frac{1}{4\mu^{2}\left(\zeta\right)}\left\{ \hat{P}+4\left[\mu\left(\zeta\right)\frac{d\mu\left(\zeta\right)}{d\zeta}+\phi\left(\zeta\right)\right]\hat{Q}-\frac{m^{2}}{r^{2}}\right\} \nonumber \\
 & + & r^{2}\left[\frac{d\phi\left(\zeta\right)}{d\zeta}-\frac{\phi^{2}\left(\zeta\right)}{\mu^{2}\left(\zeta\right)}-\frac{2\phi\left(\zeta\right)}{\mu\left(\zeta\right)}\frac{d\mu\left(\zeta\right)}{d\zeta}\right],
\end{eqnarray}
where the following two operators have been introduced: 
\begin{equation}
\hat{P}=\frac{\partial^{2}}{\partial r^{2}}+\frac{1}{r}\frac{\partial}{\partial r};\quad\hat{Q}=i\left(r\frac{\partial}{\partial r}+1\right).
\end{equation}

As will be shown further on, a simple characterization of the cylindrical
paraxial OAM beams can be derived from the definitions:
\begin{equation}
a=\mu\frac{d\mu}{d\zeta}+\phi;\quad b=\mu^{2}\left[\frac{d\phi}{d\zeta}-\frac{\phi^{2}}{\mu^{2}}-\frac{2\phi}{\mu}\frac{d\mu}{d\zeta}\right],\label{eq:a=000026b}
\end{equation}
which allow to rewrite equation (\ref{eq:SchrodingerEquation}) as:
\begin{equation}
i\frac{\partial u\left(r,\zeta\right)}{\partial\zeta}=\left\{ \frac{1}{4\mu^{2}\left(\zeta\right)}\left[\hat{P}+4a\hat{Q}-\frac{m^{2}}{r^{2}}\right]+\frac{br^{2}}{\mu^{2}\left(\zeta\right)}\right\} u\left(r,\zeta\right).\label{eq:parametricSchrodingerEquation}
\end{equation}
It is interesting to note that both $\hat{P}$ and $\hat{Q}$ are
Hermitian operators with respect to the integration measure $rdr$,
therefore the Hamiltonian $\hat{H}\left(\zeta\right)$ (term in braces
in the above expression) is Hermitian if $a,\:b\in\mathbb{R}$ and
$\mu^{2}\left(\zeta\right)$ corresponds to a real valued function
of the $\zeta$-coordinate: in this particular case, complete sets
of orthogonal solutions are obtained for the stationary Schr\"{o}dinger
equation. These and other classes of solutions are derived from the
choice of the parameters $a$, $b$ and by establishing suitable functions
$\mu\left(\zeta\right)$, $\phi\left(\zeta\right)$ which satisfy
(\ref{eq:a=000026b}).

\subsection{Bessel beams\label{sec:Bessel-beams}}

With the choice $a=b=0$, $\mu\left(\zeta\right)=1$ and $\phi\left(\zeta\right)=0$,
equation (\ref{eq:parametricSchrodingerEquation}) reads:
\begin{equation}
i\frac{\partial u_{B}\left(r,\zeta\right)}{\partial\zeta}=\frac{1}{4}\left[\frac{\partial^{2}}{\partial r^{2}}+\frac{1}{r}\frac{\partial}{\partial r}-\frac{m^{2}}{r^{2}}\right]u_{B}\left(r,\zeta\right),
\end{equation}
where the subscript $B$ has been introduced; alternatively, going
back to the original coordinate system: 
\begin{equation}
ik\frac{\partial u_{B}\left(\rho,z\right)}{\partial z}=\frac{1}{2}\left[\frac{\partial^{2}}{\partial\rho^{2}}+\frac{1}{\rho}\frac{\partial}{\partial\rho}-\frac{m^{2}}{\rho^{2}}\right]u_{B}\left(\rho,z\right).\label{eq:equationForBessel}
\end{equation}
For a paraxial beam of the form $\psi_{B}\left(\rho,\varphi,z\right)=g_{B}\left(\rho,\varphi,z\right)\exp\left(-ik_{z}z\right)$,
the following relation holds: 
\begin{equation}
k_{z}=\sqrt{k^{2}-k_{x}^{2}-k_{y}^{2}}\approx k-\frac{k_{\rho}^{2}}{2k},
\end{equation}
where $k_{\rho}^{2}=k_{x}^{2}+k_{y}^{2}$. Then we can write $\psi_{B}\left(\rho,\varphi,z\right)=f_{B}\left(\rho,\varphi,z\right)\exp\left(-ikz\right)$
with $f_{B}\left(\rho,\varphi,z\right)=g_{B}\left(\rho,\varphi,z\right)\exp\left(ik_{\rho}^{2}z/2k\right)$.
Taking into account expression (\ref{eq:Ansatz}), we infer:
\begin{equation}
u_{B}\left(\rho,z\right)\propto B\left(\rho\right)\exp\left(i\frac{k_{\rho}^{2}z}{2k}\right),
\end{equation}
which, substituted in equation (\ref{eq:equationForBessel}), gives:
\begin{equation}
\rho^{2}\frac{d^{2}B\left(\rho\right)}{d\rho^{2}}+\rho\frac{dB\left(\rho\right)}{d\rho}+\left(\rho^{2}k_{\rho}^{2}-m^{2}\right)B\left(\rho\right)=0.\label{eq:Bessel}
\end{equation}
Equation (\ref{eq:Bessel}) corresponds exactly to (\ref{eq:BesselEquation})
and, as mentioned above, its regular solution $B\left(\rho\right)$
is provided by the Bessel function of the first kind $J_{\left|m\right|}\left(k_{\rho}\rho\right)$,
therefore leading to:
\begin{equation}
f_{m}^{B}\left(\rho,\varphi,z;k_{\rho}\right)=C_{m}^{B}J_{\left|m\right|}\left(k_{\rho}\rho\right)\exp\left(i\frac{k_{\rho}^{2}z}{2k}+im\varphi\right).
\end{equation}

The paraxial modes $\Psi_{m}^{B}\left(\rho,\varphi,z,t;k_{\rho}\right)=f_{m}^{B}\left(\rho,\varphi,z;k_{\rho}\right)\mathrm{e}^{-ikz+i\omega t}$
are nothing but the above discussed BBs (\ref{eq:BesselModes}). By
virtue of their completeness and of the fact that they satisfy both
the exact and the paraxial wave equations, BBs are well suited for
implementing non-paraxial extensions of known beams families which
belong originally to the set of paraxial solutions \cite{Andrews2013}. 

\subsection{Laguerre-Gaussian beams}

\begin{figure}
\noindent \begin{centering}
\includegraphics[width=0.8\textwidth]{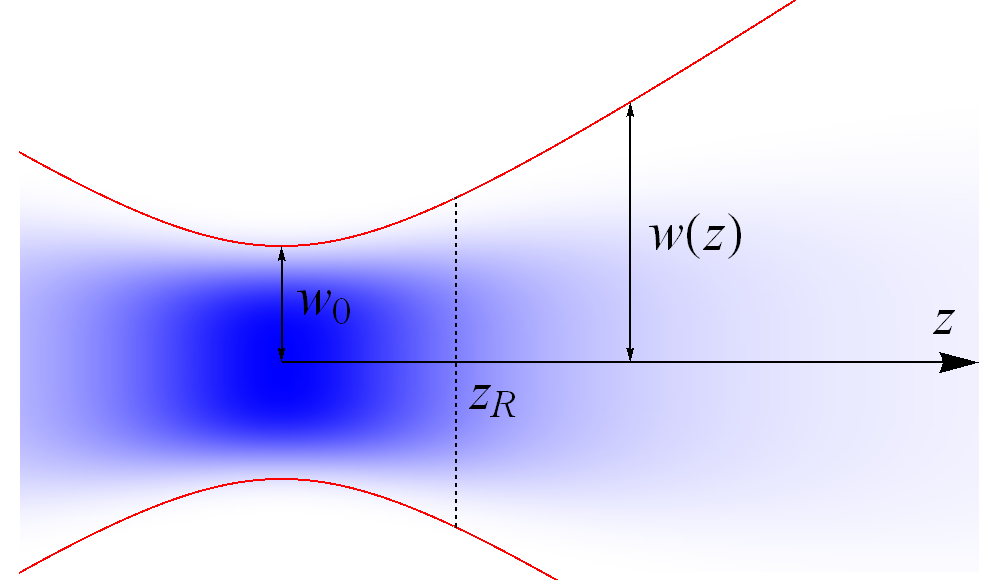}
\par\end{centering}
\caption{Propagation of the fundamental Gaussian wave $\psi_{00}^{LG}$: longitudinal
intensity profile with some beam parameters.\label{fig:GaussPropag}}
\end{figure}

Let now $a=0$, $b=-1$, $\mu\left(\zeta\right)=\sqrt{1+\zeta^{2}}$
and $\phi\left(\zeta\right)=-\zeta$:
\begin{equation}
i\frac{\partial u_{LG}\left(r,\zeta\right)}{\partial\zeta}=\frac{1}{4\mu^{2}\left(\zeta\right)}\left[\frac{\partial^{2}}{\partial r^{2}}+\frac{1}{r}\frac{\partial}{\partial r}-\frac{m^{2}}{r^{2}}-4r^{2}\right]u_{LG}\left(r,\zeta\right).\label{eq:EquationForLG}
\end{equation}
In order to solve equation (\ref{eq:EquationForLG}), it is convenient
to start from the following assumption:
\begin{equation}
u_{LG}\left(r,\zeta\right)\propto r^{\left|m\right|}LG\left(2r^{2}\right)\exp\left[-r^{2}-i\Theta\left(\zeta\right)\right],
\end{equation}
where $LG\left(2r^{2}\right)$ and $\Theta\left(\zeta\right)$ are
functions to be determined. After some straightforward algebraic steps,
equation (\ref{eq:EquationForLG}) reduces to:
\begin{equation}
4r^{\left|m\right|+2}\left(LG^{''}-LG^{'}\right)+r^{\left|m\right|}\left\{ 2\left(\left|m\right|+1\right)LG^{'}-\left[\left|m\right|+1+\mu^{2}\frac{d\Theta}{d\zeta}\right]LG\right\} =0,\label{eq:searchLaguerre}
\end{equation}
with the definitions $LG^{'}=dLG\left(x\right)/dx$ and $LG^{''}=d^{2}LG\left(x\right)/dx^{2}$.
Since an equation which holds for every $\zeta$ is needed, the coefficient
with $\zeta$-dependence must be set equal to a constant, a conventional
choice being $-\left(2p+\left|m\right|+1\right)$, with $p\in\mathbb{N}$:
\begin{equation}
\left|m\right|+1+\mu^{2}\left(\zeta\right)\frac{d\Theta\left(\zeta\right)}{d\zeta}=-2p.
\end{equation}
The equation is easily solved and gives:
\begin{equation}
\Theta\left(\zeta\right)=-\left(2p+\left|m\right|+1\right)\arctan\left(\zeta\right).
\end{equation}
Hence, expression (\ref{eq:searchLaguerre}) becomes:
\begin{equation}
2r^{2}LG^{''}\left(2r^{2}\right)+\left(\left|m\right|+1-2r^{2}\right)LG^{'}\left(2r^{2}\right)+pLG\left(2r^{2}\right),
\end{equation}
whose regular solution is the generalized Laguerre polynomial $L_{p}^{\left|m\right|}\left(2r^{2}\right)$
\cite{Abramowitz1972} and leads to:
\begin{eqnarray}
f_{pm}^{LG}\left(\rho,\varphi,z\right) & = & \frac{C_{pm}^{LG}}{w\left(z\right)}\left[\frac{\rho}{w\left(z\right)}\right]^{\left|m\right|}L_{p}^{\left|m\right|}\left[\frac{2\rho^{2}}{w^{2}\left(z\right)}\right]\exp\left(im\varphi\right)\nonumber \\
 & \times & \exp\left[-\frac{\rho^{2}}{w_{0}^{2}\left(1-i\frac{z}{z_{R}}\right)}-i\Theta\left(\frac{z}{z_{R}}\right)\right].\label{eq:Laguerre-Gauss}
\end{eqnarray}
In expression (\ref{eq:Laguerre-Gauss}), the definition $w\left(z\right)=w_{0}\,\mu\left(z/z_{R}\right)$
has been introduced, $w_{0}$ corresponds to the waist of the Gaussian
beam, $z_{R}$ is the Rayleigh distance (see Figure \ref{fig:GaussPropag})
and $\Theta\left(\zeta\right)$ the Gouy phase \cite{Siegman1986}.
The constant term $C_{pm}^{LG}$ ensures correct dimension and normalization
of the beam profile. 

\begin{figure}
\noindent \begin{centering}
\includegraphics[width=1\textwidth]{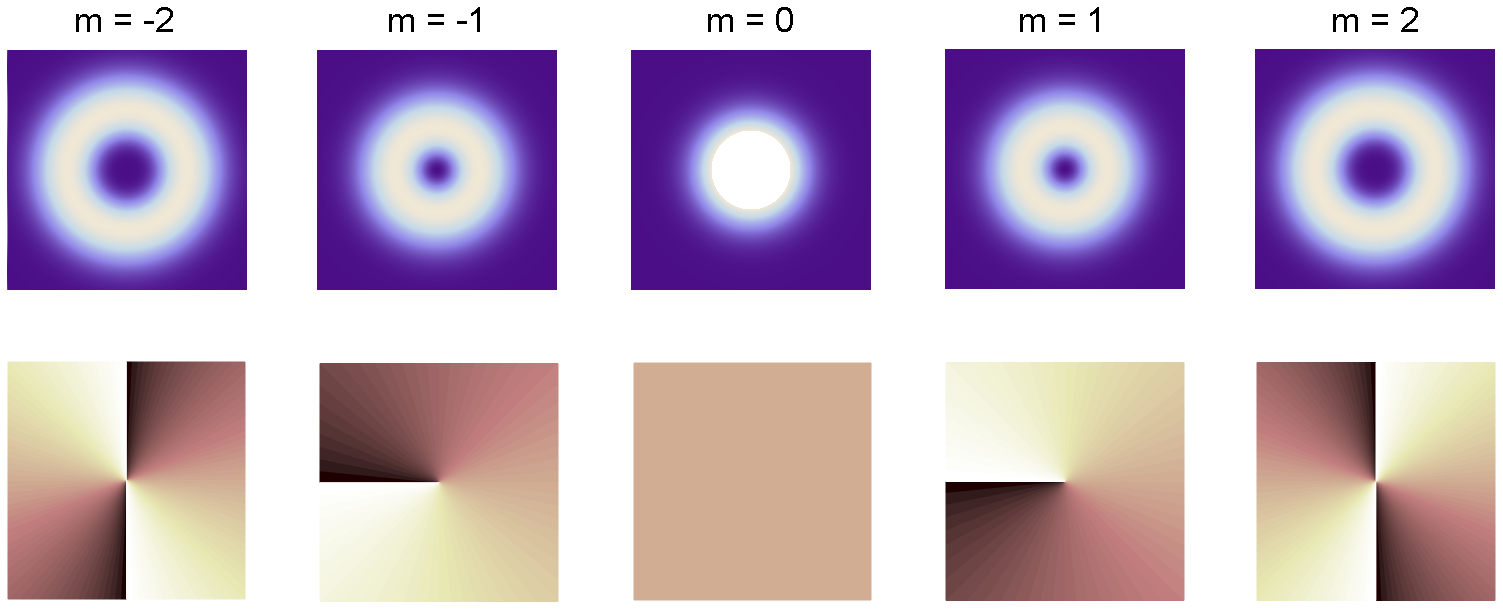}
\par\end{centering}
\caption{LG beam intensity (upper row) and phase (lower row) transverse profiles ($y$-coordinate versus $x$-coordinate) for
$p=0$.\label{fig:LGs}}
\end{figure}

The cylindrical modes $\Psi_{pm}^{LG}\left(\rho,\varphi,z,t\right)=f_{pm}^{LG}\left(\rho,\varphi,z\right)\mathrm{e}^{-ikz+i\omega t}$
are named \textit{Laguerre-Gaussian }(LG\nomenclature{LG}{Laguerre-Gaussian})\textit{
beams} and represent another complete set of orthogonal paraxial OAM
solutions of the scalar wave equation. The transverse intensity and phase profiles of a
representative LG beam are displayed in Figure \ref{fig:LGs} for different values of the topological charge $m$. Unlike BBs, the LG modes diverge
during propagation (in this regard, $z_{R}$ can be seen as the diffraction
scale), they are spatially confined and carry a finite amount of energy,
therefore can be physically realized, at least up to some degree of
accuracy. 

\begin{figure}[t]
\noindent \begin{centering}
\includegraphics[width=0.95\textwidth]{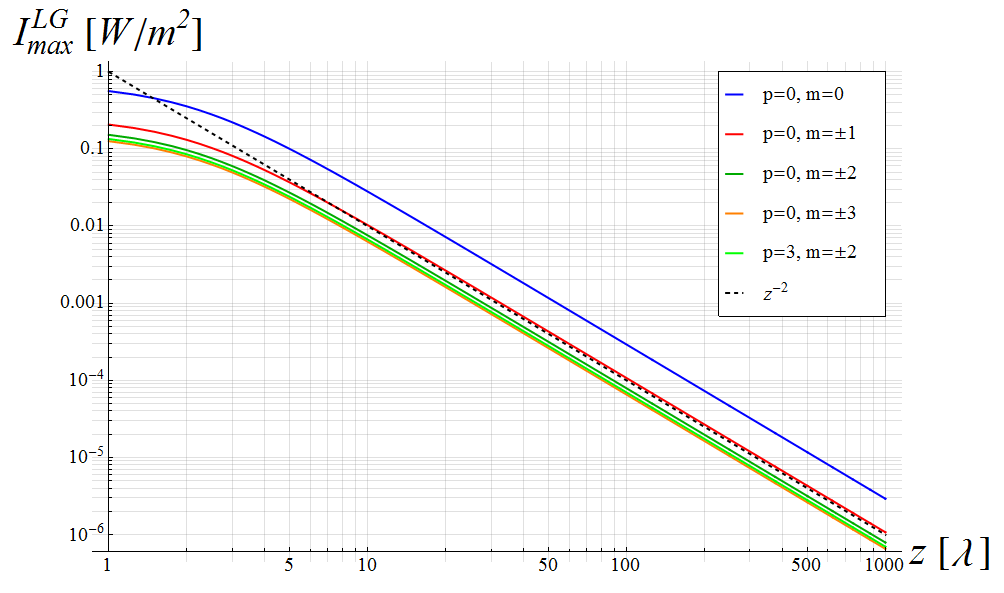}
\par\end{centering}
\caption{Evolution of the principal intensity maximum of some LG beams as a
function of the propagation distance for $w_{0}=\lambda$.\label{fig:GaussMaximum}}
\end{figure}

LG beams are shape-invariant modes: during free propagation, their
intensity profile maintains the same shape and it is just scaled owing
to diffraction and energy conservation. The radius of the primary
intensity maximum evolves with $z$ according to the law:
\begin{equation}
\rho_{max}^{LG}\left(z\right)=\rho_{max,0}^{LG}\,\sqrt{1+\left(\frac{z}{z_{R}}\right)^{2}},\label{eq:rMaxLG}
\end{equation}
where $\rho_{max,0}^{LG}=\rho_{max}^{LG}\left(0\right)$ vanishes for $m=0$. At large distances, for $m\neq0$, we infer
$\rho_{max}^{LG}\left(z\right)\propto z$. By replacing $\rho$ with
expression (\ref{eq:rMaxLG}) in the square modulus of (\ref{eq:Laguerre-Gauss}),
we easily get: 
\begin{equation}
I_{max}^{LG}\left(z\right)=\left|f_{pm}^{LG}\left[\rho_{max}^{LG}\left(z\right),\varphi,z\right]\right|^{2}=I_{max,0}^{LG}\left[1+\left(\frac{z}{z_{R}}\right)^{2}\right]^{-1},
\end{equation}
being $I_{max,0}^{LG}=I_{max}^{LG}\left(0\right)$ the maximum intensity
of the original profile. Therefore, for $z\gg z_{R}$, the asymptotic
behavior $I_{max}^{LG}\left(z\right)\propto z^{-2}$ is obtained,
as shown in Figure \ref{fig:GaussMaximum}.

\begin{figure}[h]
\noindent \begin{centering}
\includegraphics[width=0.95\textwidth]{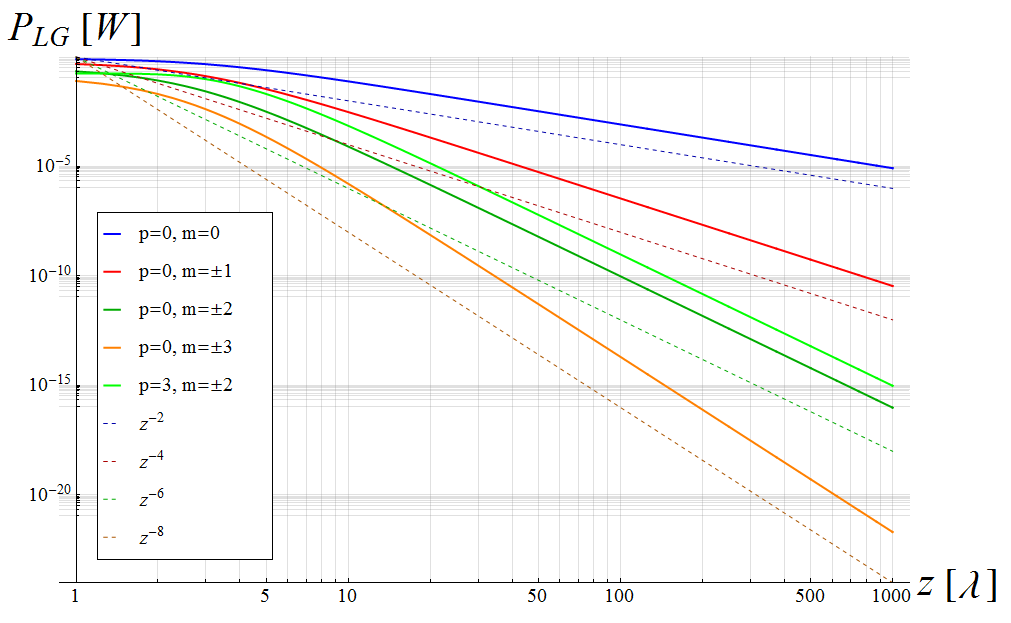}
\par\end{centering}
\caption{Evolution of the integral of the intensity of some LG beams over a
centered circular region with radius $w_{0}$ as a function of $z$,
for $w_{0}=\lambda$.\label{fig:GaussPower}}
\end{figure}

Let us now consider the decay of the intensity of a LG beam at a fixed
radial coordinate $\bar{\rho}$ (for $m=0$, this also includes the
case of the primary maximum at $\bar{\rho}=0$); the square modulus
of (\ref{eq:Laguerre-Gauss}) is given by:
\begin{equation}
I_{LG}\left(z\right)=\left|f_{pm}^{LG}\left[\bar{\rho},\varphi,z\right]\right|^{2}\propto\left[w^{2}\left(z\right)\right]^{-\left|m\right|-1}\left|L_{p}^{\left|m\right|}\left[\frac{2\bar{\rho}^{2}}{w^{2}\left(z\right)}\right]\right|^{2}\mathrm{e}^{-\frac{2\bar{\rho}^{2}}{w^{2}\left(z\right)}}.
\end{equation}
For large enough distances, the term which contains the Laguerre polynomial
tends to a constant and so does the exponential one, thus implying
an asymptotic behavior of the form $I_{LG}\left(z\right)\propto z^{-2\left|m\right|-2}.$
The same law also describes the far-field evolution of the LG beam
power collected on a central surface of limited size placed at distance
$z$, reported in Figure \ref{fig:GaussPower}.


\subsection{Hypergeometric beams}

By choosing $a=-1/2$, $b=0$, $\mu\left(\zeta\right)=\sqrt{-\zeta}$
and $\phi\left(\zeta\right)=0$, equation (\ref{eq:parametricSchrodingerEquation})
reduces to:
\begin{equation}
i\frac{\partial u_{H}\left(r,\zeta\right)}{\partial\zeta}=\frac{1}{4\mu^{2}\left(\zeta\right)}\left[\frac{\partial^{2}}{\partial r^{2}}+\frac{1}{r}\frac{\partial}{\partial r}-2i\left(r\frac{\partial}{\partial r}+1\right)-\frac{m^{2}}{r^{2}}\right]u_{H}\left(r,\zeta\right).\label{eq:equationForHypergeometric}
\end{equation}
If a field amplitude of the form:
\begin{equation}
u_{H}\left(r,\zeta\right)\propto r^{\left|m\right|}Z_{H}\left(\zeta\right)H\left(ir^{2}\right)
\end{equation}
is assumed, expression (\ref{eq:equationForHypergeometric}) becomes:
\begin{equation}
ir^{2}H^{''}+\left[\left(\left|m\right|+1\right)-ir^{2}\right]H^{'}-\left[\frac{\left|m\right|+1}{2}+\frac{\mu^{2}}{Z_{H}}\frac{dZ_{H}}{d\zeta}\right]H=0,\label{eq:searchForHypergeometric}
\end{equation}
which holds true for any $\zeta$ under the following hypothesis:
\begin{equation}
\frac{\mu^{2}\left(\zeta\right)}{Z_{H}\left(\zeta\right)}\frac{dZ_{H}\left(\zeta\right)}{d\zeta}=-\frac{i\gamma}{2},\label{eq:EquationForZ_H}
\end{equation}
where $\gamma\in\mathbb{R}$. From (\ref{eq:EquationForZ_H}), we
get:
\begin{equation}
Z_{H}\left(\zeta\right)=\sqrt{\zeta^{i\gamma}}
\end{equation}
and finally, from (\ref{eq:searchForHypergeometric}) and (\ref{eq:EquationForZ_H}):
\begin{equation}
ir^{2}H^{''}\left(ir^{2}\right)+\left[\left(\left|m\right|+1\right)-ir^{2}\right]H^{'}\left(ir^{2}\right)-\left[\frac{\left|m\right|+1}{2}-\frac{i\gamma}{2}\right]H\left(ir^{2}\right)=0.
\end{equation}
This last equation admits a regular solution, given in terms of the
Kummer hypergeometric confluent function $_{1}F_{1}\left[\left(\left|m\right|+1-i\gamma\right)/2,\left|m\right|+1;\,ir^{2}\right]$
(see, for example, \cite{Abramowitz1972}), which allows to obtain:
\begin{eqnarray}
f_{m}^{H}\left(\rho,\varphi,z;\gamma\right) & = & C_{m}^{H}\left(\gamma\right)\left[\frac{k\rho^{2}}{2z}\right]^{\frac{\left|m\right|}{2}}\left(\frac{z}{z_{R}}\right)^{\frac{i\gamma-1}{2}}\exp\left(im\varphi\right)\nonumber \\
 & \times & _{1}F_{1}\left(\frac{\left|m\right|+1}{2}-\frac{i\gamma}{2},\left|m\right|+1;\,-i\frac{k\rho^{2}}{2z}\right),\label{eq:HypergeometricModes}
\end{eqnarray}
where the dimensional constant $C_{m}^{H}\left(\gamma\right)$ is
a function of the beam parameters and can be fixed by imposing the
orthogonality condition.

\begin{figure}[t]
\noindent \begin{centering}
\includegraphics[width=1\textwidth]{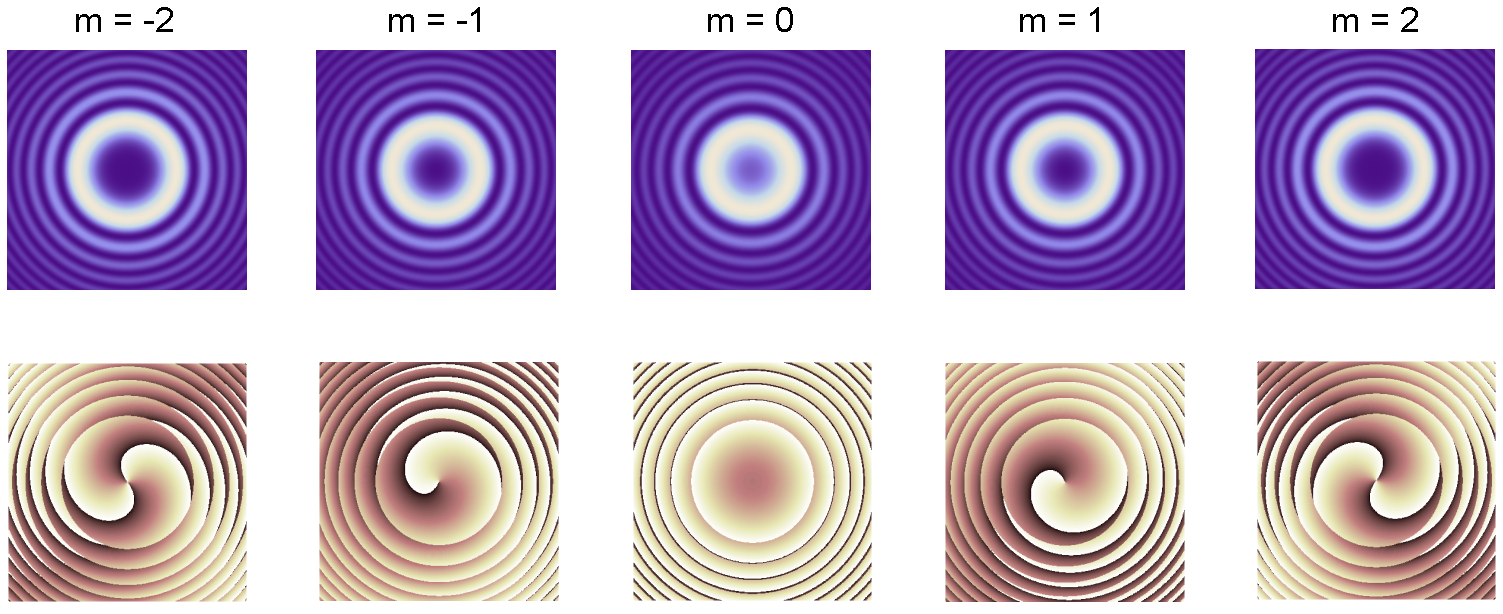}
\par\end{centering}
\caption{HyG beam intensity (upper row) and phase (lower row) transverse profiles ($y$-coordinate versus $x$-coordinate) for
$\gamma=1$.\label{fig:HyGs}}
\end{figure}

The \textit{hypergeometric }(HyG\nomenclature{HyG}{hypergeometric})\textit{
beams} $\Psi_{m}^{H}\left(\rho,\varphi,z,t;\gamma\right)=f_{m}^{H}\left(\rho,\varphi,z;\gamma\right)\mathrm{e}^{-ikz+i\omega t}$
constitute the third complete family of orthogonal OAM solutions of
the scalar wave equation in the paraxial regime; unfortunately, like
BBs, they carry infinite energy and are not square-integrable \cite{Kotlyar2007}. The transverse intensity and phase profiles of a
representative HyG beam are displayed in Figure \ref{fig:HyGs} for different values of the topological charge $m$.
Equation (\ref{eq:HypergeometricModes}) differs from the standard
formula found in literature by a minus sign in the argument of the
Kummer function, which is due to the choice of a propagation term
of the form $\exp\left(-ikz+i\omega t\right)$ instead of $\exp\left(ikz-i\omega t\right)$,
showing the two cases an opposite sign in front of the $z$-derivative
in the corresponding versions of the paraxial Helmholtz equation. 

\subsection{Hypergeometric-Gaussian beams}

Taking into consideration the possibility of using complex parameters,
we can set $a=i/2$, $b=0$, $\mu\left(\zeta\right)=\sqrt{\zeta\left(\zeta+i\right)}$
and $\phi\left(\zeta\right)=-\zeta$:
\begin{equation}
i\frac{\partial u_{HG}\left(r,\zeta\right)}{\partial\zeta}=\frac{1}{4\mu^{2}\left(\zeta\right)}\left[\frac{\partial^{2}}{\partial r^{2}}+\frac{1}{r}\frac{\partial}{\partial r}-2\left(r\frac{\partial}{\partial r}+1\right)-\frac{m^{2}}{r^{2}}\right]u_{HG}\left(r,\zeta\right).\label{eq:EquationForHyperGauss}
\end{equation}
Let us suppose the function $u_{HG}$ to be factorized as:
\begin{equation}
u_{HG}\left(r,\zeta\right)\propto r^{\left|m\right|}Z_{HG}\left(\zeta\right)HG\left(r^{2}\right),
\end{equation}
then (\ref{eq:EquationForHyperGauss}) reduces to:
\begin{equation}
r^{2}HG^{''}+\left[\left(\left|m\right|+1\right)-r^{2}\right]HG^{'}-\left[\frac{\left|m\right|+1}{2}+i\frac{\mu^{2}}{Z_{HG}}\frac{dZ_{HG}}{d\zeta}\right]HG=0.\label{eq:SearchForHyperGauss}
\end{equation}
Following the usual procedure, it is required that:
\begin{equation}
i\frac{\mu^{2}\left(\zeta\right)}{Z_{HG}\left(\zeta\right)}\frac{dZ_{HG}\left(\zeta\right)}{d\zeta}=-\frac{\eta+\left|m\right|+1}{2},
\end{equation}
where $\eta\in\mathbb{R}$. A quick integration gives:
\begin{equation}
Z_{HG}\left(\zeta\right)=\left(\frac{\zeta}{\zeta+i}\right)^{\frac{\eta+\left|m\right|+1}{2}},\label{eq:Z_HG}
\end{equation}
whereas the Kummer equation \cite{Abramowitz1972} arises from (\ref{eq:SearchForHyperGauss}):
\begin{equation}
r^{2}HG^{''}\left(r^{2}\right)+\left[\left(\left|m\right|+1\right)-r^{2}\right]HG^{'}\left(r^{2}\right)+\frac{\eta}{2}HG\left(r^{2}\right)=0,
\end{equation}
\begin{equation}
HG\left(r^{2}\right)={}_{1}F_{1}\left[-\frac{\eta}{2},\left|m\right|+1;\,r^{2}\right].\label{eq:Kummer}
\end{equation}
Expression (\ref{eq:Ansatz}) leads to:
\begin{eqnarray}
f_{m}^{HG}\left(\rho,\varphi,z;\eta\right) & = & C_{m}^{HG}\left(\eta\right)\left(\frac{\rho}{w_{0}}\right)^{\left|m\right|}\left(\frac{z}{z_{R}}\right)^{\frac{\eta}{2}}\left(\frac{z}{z_{R}}+i\right)^{-\left(1+\left|m\right|+\frac{\eta}{2}\right)}\exp\left(im\varphi\right)\nonumber \\
 & \times & _{1}F_{1}\left[-\frac{\eta}{2},\left|m\right|+1;\,\frac{k\rho^{2}}{2z\left(\frac{z}{z_{R}}+i\right)}\right]\exp\left[-i\frac{\rho^{2}}{w_{0}^{2}\left(\frac{z}{z_{R}}+i\right)}\right].
\end{eqnarray}

The modes represented by $\Psi_{m}^{HG}\left(\rho,\varphi,z,t;\eta\right)=f_{m}^{HG}\left(\rho,\varphi,z;\eta\right)\mathrm{e}^{-ikz+i\omega t}$
constitute the family of the \textit{hypergeometric-Gaussian} (HyGG\nomenclature{HyGG}{hypergeometric-Gaussian})\textit{
beams}, an overcomplete set of non-orthogonal paraxial OAM solutions
of the scalar wave equation \cite{Karimi2007}. It can be proven that
these configurations carry a finite power (and therefore can be experimentally
approximated) as long as the condition $\eta\geq-\left|m\right|$
is satisfied. 

The HyGG beams are not shape-invariant modes under free propagation;
as far as the asymptotic behavior is concerned, the radius of the
primary maximum of their intensity profile follows the same law as
the LG case for $m\neq0$, namely $\rho_{max}^{HG}\left(z\right)\propto z$,
which implies the usual decay of the intensity maximum at large distances,
$I_{max}^{HG}\left(z\right)=\left|f_{m}^{HG}\left[\rho_{max}^{HG}\left(z\right),\varphi,z;\eta\right]\right|^{2}\propto z^{-2}$.
Also the asymptotic evolution of the intensity at fixed $\rho=\bar{\rho}$
is found to be $I_{HG}\left(z\right)\propto z^{-2\left|m\right|-2}$,
as for the LG beams.

\subsection{Hypergeometric-Gaussian beams of the second kind}

Another interesting family of cylindrical modes is found by choosing
$a=-i/2$, $b=0$, $\mu\left(\zeta\right)=\sqrt{1-i\zeta}$ and $\phi\left(\zeta\right)=0$.
In this case, equation (\ref{eq:parametricSchrodingerEquation}) reads:
\begin{equation}
i\frac{\partial u_{HG_{II}}}{\partial\zeta}=\frac{1}{4\mu^{2}}\left[\frac{\partial^{2}}{\partial r^{2}}+\frac{1}{r}\frac{\partial}{\partial r}+2\left(r\frac{\partial}{\partial r}+1\right)-\frac{m^{2}}{r^{2}}\right]u_{HG_{II}}.
\end{equation}
Let us suppose:
\begin{equation}
u_{HG_{II}}\left(r,\zeta\right)\propto r^{\left|m\right|}Z_{HG_{II}}\left(\zeta\right)\exp\left(-r^{2}\right)HG_{II}\left(r^{2}\right),
\end{equation}
which implies: 
\begin{equation}
r^{2}HG_{II}^{''}+\left[\left(\left|m\right|+1\right)-r^{2}\right]HG_{II}^{'}-\left[\frac{\left|m\right|+1}{2}+i\frac{\mu^{2}}{Z_{HG_{II}}}\frac{dZ_{HG_{II}}}{d\zeta}\right]HG_{II}=0.\label{eq:SearchForHyperGaussII}
\end{equation}
Let now:
\begin{equation}
i\frac{\mu^{2}\left(\zeta\right)}{Z_{HG_{II}}\left(\zeta\right)}\frac{dZ_{HG_{II}}\left(\zeta\right)}{d\zeta}=-\frac{\eta+\left|m\right|+1}{2},
\end{equation}
whose solution is given by:
\begin{equation}
Z_{HG_{II}}\left(\zeta\right)=\left(1-i\zeta\right)^{-\frac{\eta+\left|m\right|+1}{2}}.
\end{equation}
Equation (\ref{eq:SearchForHyperGaussII}) has the same form as (\ref{eq:SearchForHyperGauss}),
thus $HG_{II}\left(r^{2}\right)=HG\left(r^{2}\right)$, expression
(\ref{eq:Kummer}). Finally:
\begin{eqnarray}
f_{m}^{HG_{II}}\left(\rho,\varphi,z;\eta\right) & \!\!=\!\! & C_{m}^{HG_{II}}\left(\eta\right)\left(\frac{\rho}{w_{0}}\right)^{\left|m\right|}\left(1-i\frac{z}{z_{R}}\right)^{-\left(1+\left|m\right|+\frac{\eta}{2}\right)}\exp\left(im\varphi\right)\nonumber \\
 & \!\!\times\!\! & _{1}F_{1}\left[-\frac{\eta}{2},\left|m\right|+1;\,\frac{\rho^{2}}{w_{0}^{2}\left(1-i\frac{z}{z_{R}}\right)}\right]\exp\left[-\frac{\rho^{2}}{w_{0}^{2}\left(1-i\frac{z}{z_{R}}\right)}\right].\label{eq:Hypergeometric-GaussianII}
\end{eqnarray}

Paraxial modes of the kind $\varPsi_{m}^{HG_{II}}\left(\rho,\varphi,z,t;\eta\right)=f_{m}^{HG_{II}}\left(\rho,\varphi,z;\eta\right)\mathrm{e}^{-ikz+i\omega t}$
are known as \textit{hypergeometric-Gaussian type-II }(HyGG-II\nomenclature{HyGG-II}{hypergeometric-Gaussian type-II})\textit{
beams} and, like the HyGG modes, represent an overcomplete set of
non-orthogonal OAM solutions of the Helmholtz equation which are square-integrable
for $\eta\geq-\left|m\right|$, condition under which they carry a
finite amount of energy \cite{Karimi2008}. 

\begin{figure}[t]
\noindent \begin{centering}
\includegraphics[width=1\textwidth]{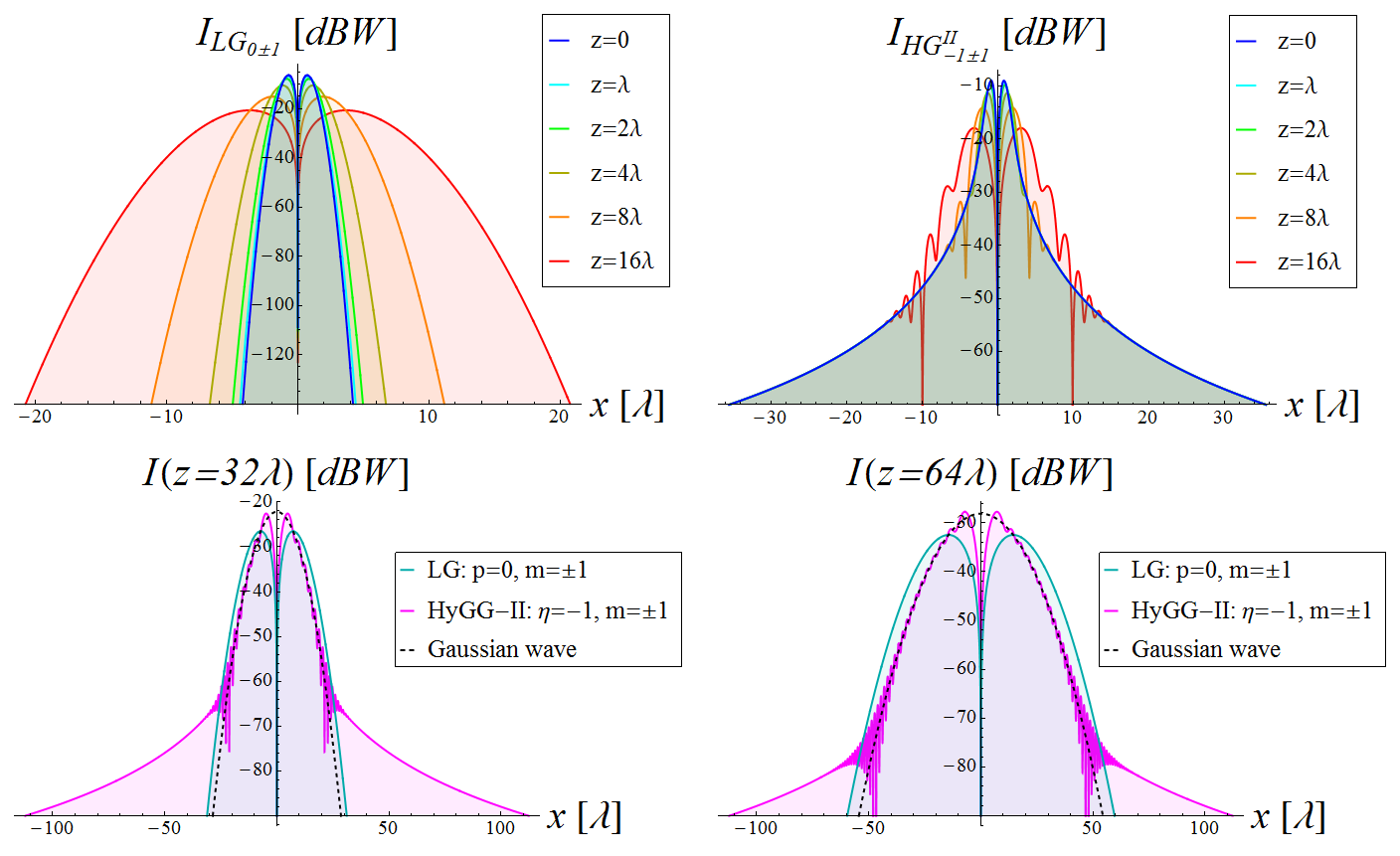}
\par\end{centering}
\caption{Comparison between the power profiles (per unit square meter) of a
LG beam and of a HyGG-II beam evaluated in $y=0$, for $w_{0}=\lambda$.\label{fig:LGHGcomparison}}
\end{figure}

The HyGG-II beams are not shape-invariant modes under free propagation
(see Figure \ref{fig:LGHGcomparison}). Following the standard convention,
where a propagation term of the form $\exp\left(ikz-i\omega t\right)$
is considered, inducing a change of sign in the $z$-derivative in
(\ref{eq:paraxialHelmholtzEquation}), the function $\mu\left(\zeta\right)$
has the form $\sqrt{1+i\zeta}$ and equation (\ref{eq:Hypergeometric-GaussianII})
changes accordingly.

Despite their name, resembling the previous family, the HyGG-II beams
show significant differences with respect to the other known solutions.
First, the radius of their primary intensity maximum increases with
$z$ according to two possible asymptotic behaviors, depending on
the choice of the $\eta$ and $m$ indices: when $\eta=-\left|m\right|\neq0$,
a square root law is found, $\rho_{max}^{HG_{II}}\left(z\right)\propto\sqrt{z}$,
whereas the usual linear behavior $\rho_{max}^{HG_{II}}\left(z\right)\propto z$
is quickly reached by all the configurations in which $\eta>0$ for
every $m$ or $\eta\geq0$ for $m\neq0$ (the doughnut-shaped asymptotic
profile is always present, except for the standard Gaussian wave $\eta=m=0$).
It is fundamental to understand that the square root divergence of
the primary intensity maximum relative to the HyGG-II modes in the
first subclass does not apply to the beam profile in its entirety,
as can be inferred from the comparison reported in Figure \ref{fig:LGHGcomparison}. 

Since the $\eta$ parameter must be greater than or equal to $-\left|m\right|$
in order to ensure square-integrability, the whole range $-\left|m\right|<\eta<0$
where $m$ is not zero can be explored: in this region, as a consequence
of the very rippled and variable field pattern, the primary intensity
maximum oscillates at different radii in the beam profile and 
it evolves alternating between the two possible above described laws
as a function of the propagation coordinate, asymptotically stabilizing 
to the linear evolution in $z$ (see Figure \ref{fig:radiusHGII}).

In order to establish the decay law of the primary intensity maximum,
the two asymptotic cases must be considered separately. If $\eta=-\left|m\right|\neq0$,
the argument of the Kummer function in (\ref{eq:Hypergeometric-GaussianII}),
as well as that of the exponential, tends to a constant value for
large $z$ and so we immediately get $I_{max}^{HG_{II}}\left(z\right)=\left|f_{m}^{HG_{II}}\left[\rho_{max}^{HG_{II}}\left(z\right),\varphi,z;-\left|m\right|\right]\right|^{2}\propto z^{-2}$.
On the other hand, when $\eta>-\left|m\right|$ a useful asymptotic
expansion of the Kummer function can be employed \cite{Abramowitz1972}:
\begin{figure}[t]
\noindent \begin{centering}
\includegraphics[width=0.87\textwidth]{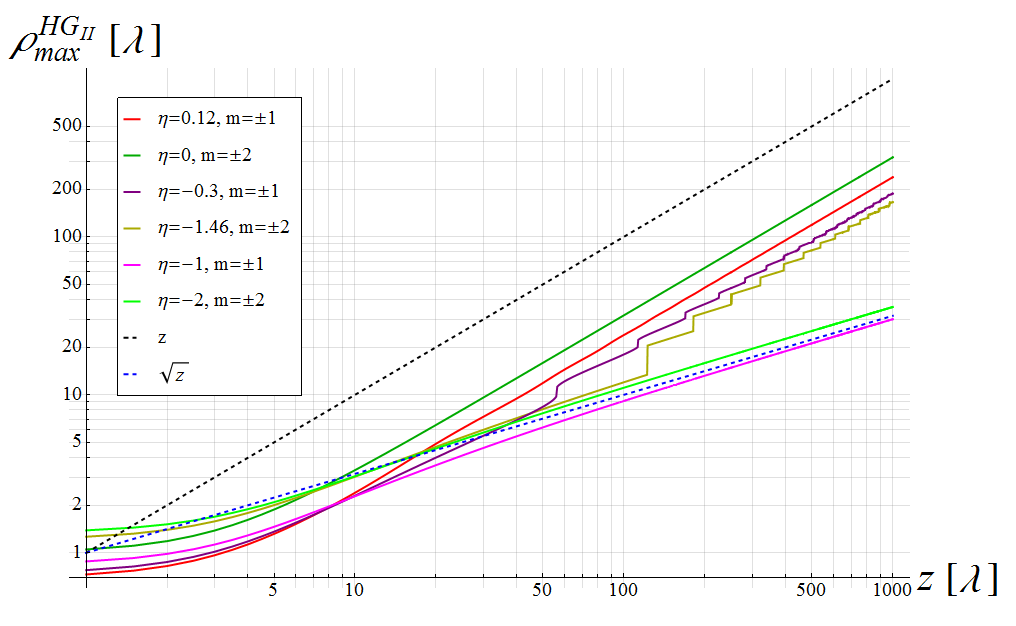}
\par\end{centering}
\caption{Evolution of the radial coordinate of the principal intensity maximum
of some HyGG-II beams as a function of $z$, for $w_{0}=\lambda$. Note that for 
$-\left|m\right|<\eta<0$ unstable maxima of the intensity profile with different 
evolution in $z$ are competing in a certain range of the propagation coordinate; 
since the plot only captures the principal maximum at each $z$, this results 
in a stair-like behavior. \label{fig:radiusHGII}}
\end{figure}
\begin{figure}[H]
\noindent \begin{centering}
\includegraphics[width=0.87\textwidth]{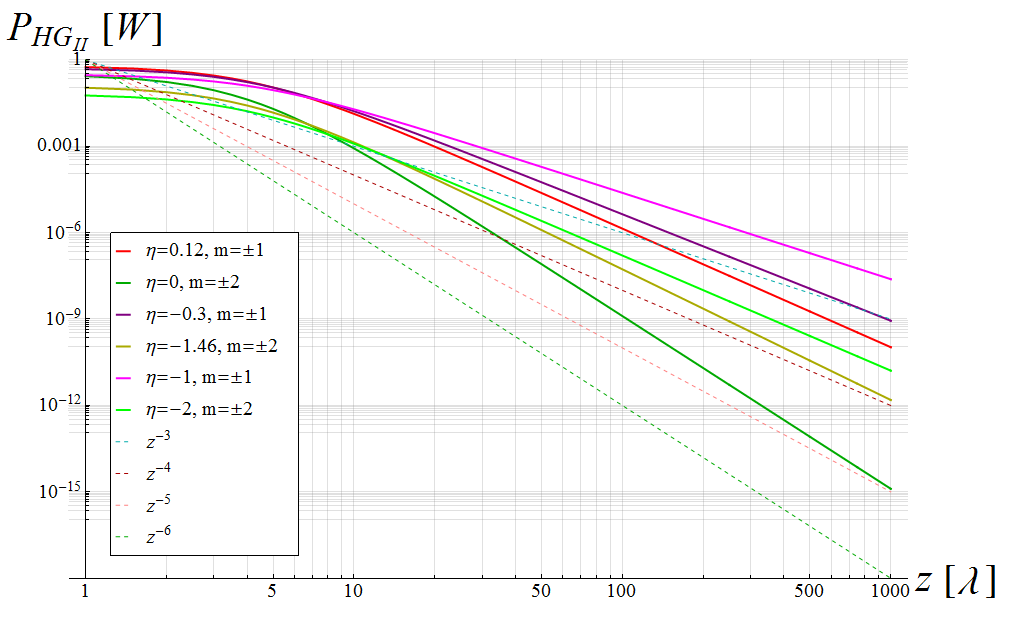}
\par\end{centering}
\caption{Evolution of the integral of the intensity of some HyGG-II beams over
a centered circular region with radius $w_{0}$ as a function of $z$,
for $w_{0}=\lambda$.\label{fig:HyperPower}}
\end{figure}
\begin{equation}
_{1}F_{1}\left(a,b;\,x\right)\approx\Gamma\left(b\right)\left[\frac{\exp\left(x\right)x^{a-b}}{\Gamma\left(a\right)}+\frac{\left(-x\right)^{-a}}{\Gamma\left(b-a\right)}\right],\label{eq:KummerAsintotics}
\end{equation}
which applies for large $\left|x\right|$. Remembering that now $\rho_{max}^{HG_{II}}\left(z\right)\propto z$,
the local far-field intensity reduces to:
\begin{equation}
I_{max}^{HG_{II}}\left(z\right)\propto z^{2\left|m\right|}z^{-\left(2+2\left|m\right|+\eta\right)}\left|\frac{\mathrm{e}^{i\pi\frac{z}{\lambda}}\left(i\pi\frac{z}{\lambda}\right)^{-\frac{\eta}{2}-\left|m\right|-1}}{\Gamma\left(-\frac{\eta}{2}\right)}+\frac{\left(-i\pi\frac{z}{\lambda}\right)^{\frac{\eta}{2}}}{\Gamma\left(\left|m\right|+1+\frac{\eta}{2}\right)}\right|^{2},
\end{equation}
where expression (\ref{eq:KummerAsintotics}) was used. Since the
second term in the sum under square modulus dominates over the first
at large $z$, this enables us to obtain $I_{max}^{HG_{II}}\left(z\right)\propto z^{-2}$,
which then holds true for every $\eta\geq-\left|m\right|$.

Finally, at fixed radial coordinate and large enough $z$, it is easy
to prove that $I_{HG_{II}}\left(z\right)\propto z^{-\eta-2\left|m\right|-2}$,
which also describes the power decay in a central region of the beam
as a function of the propagation distance, displayed in Figure \ref{fig:HyperPower}. Of special interest is the case $\eta=-\left|m\right|$, corresponding
to the best possible evolution, $z^{-\left|m\right|-2}$. 

\subsection{Bessel-Gaussian beams}

In Section \ref{sec:Bessel-beams} it was said that BBs are not square-integrable;
however, there exist paraxial solutions of the Helmholtz equation
which consists in non-diffracting modes modulated by a Gaussian envelope.
Such beams are known as \textit{Helmholtz-Gaussian }(HzG\nomenclature{HzG}{Helmholtz-Gaussian})\textit{
waves} \cite{GutierrezVega2005}. Although these modulated solutions
lose the divergence-free behavior, they present the advantage of being
square-integrable and therefore admit physical realization. In the
case of a Bessel configuration, the corresponding HzG solutions are
given by the \textit{Bessel-Gaussian }(BG\nomenclature{BG}{Bessel-Gaussian})\textit{
beams} $\varPsi_{m}^{BG}\left(\rho,\varphi,z,t;k_{\rho}\right)=f_{m}^{BG}\left(\rho,\varphi,z;k_{\rho}\right)\mathrm{e}^{-ikz+i\omega t}$,
achievable from expressions (\ref{eq:Ansatz}) and (\ref{eq:parametricSchrodingerEquation})
through the choice $a=-i$, $b=1$, $\mu\left(\zeta\right)=1-i\zeta$
and $\phi\left(\zeta\right)=\zeta$, which leads to: 
\begin{equation}
i\frac{\partial u_{BG}}{\partial\zeta}=\frac{1}{4\mu^{2}}\left[\frac{\partial^{2}}{\partial r^{2}}+\frac{1}{r}\frac{\partial}{\partial r}+4\left(r\frac{\partial}{\partial r}+1\right)-\frac{m^{2}}{r^{2}}+4r^{2}\right]u_{BG}.\label{eq:equationForBesselGauss}
\end{equation}
With reference to Section \ref{sec:Bessel-beams}:
\begin{equation}
u_{BG}\left(r,\zeta\right)\propto\exp\left(-r^{2}\right)BG\left(r\right)\exp\left(i\frac{w_{0}^{2}k_{\rho}^{2}\zeta}{4\mu\left(\zeta\right)}\right).
\end{equation}
Equation (\ref{eq:equationForBesselGauss}) is easily reduced to the
following:
\begin{equation}
r^{2}\frac{d^{2}BG\left(r\right)}{dr^{2}}+r\frac{dBG\left(r\right)}{dr}+\left(r^{2}w{}_{0}^{2}k_{\rho}^{2}-m^{2}\right)BG\left(r\right)=0,
\end{equation}
corresponding to Bessel equation in the variable $\left(\rho/\mu\right)$.
Then, from (\ref{eq:Ansatz}):
\begin{eqnarray}
f_{m}^{BG}\left(\rho,\varphi,z;k_{\rho}\right) & = & \frac{C_{m}^{BG}\left(k_{\rho}\right)}{1-i\frac{z}{z_{R}}}J_{\left|m\right|}\left(\frac{k_{\rho}\rho}{1-i\frac{z}{z_{R}}}\right)\exp\left(im\varphi\right)\nonumber \\
 & \times & \exp\left[-\frac{\rho^{2}}{w_{0}^{2}\left(1-i\frac{z}{z_{R}}\right)}+i\frac{k_{\rho}^{2}z}{2k\left(1-i\frac{z}{z_{R}}\right)}\right].\label{eq:Bessel-Gauss}
\end{eqnarray}

Once again, due to the convention employed for the propagation term,
equation (\ref{eq:Bessel-Gauss}) is not of the form usually reported
in the literature: in order to obtain the standard expression, the
whole procedure presented in Section \ref{sec:Paraxial} should be
repeated starting from the assumption $\psi\left(\rho,\varphi,z\right)=f\left(\rho,\varphi,z\right)\mathrm{e}^{ikz}$. 

The intensity distribution of the BG beams is strongly related to
the value of the transverse momentum $k_{\rho}$ as well as to the
beam waist $w_{0}$ of the Gaussian envelope \cite{Gori1987}. The
evolution of the radius of the maximum intensity follows the asymptotic
law $\rho_{max}^{BG}\left(z\right)\propto z$ (for certain choices
of the $k_{\rho}$ parameter a doughnut-shaped profile is also generated
when $m=0$, instead of the usual maximum at $\rho=0$). At large
distances, $I_{max}^{BG}\left(z\right)=\left|f_{m}^{BG}\left[\rho_{max}^{BG}\left(z\right),\varphi,z;k_{\rho}\right]\right|^{2}\propto z^{-2}$.
By exploiting the well-known relation \cite{Abramowitz1972}:
\begin{equation}
J_{\alpha}(x)\approx\frac{1}{\Gamma\left(\alpha+1\right)}\left(\frac{x}{2}\right)^{\alpha},
\end{equation}
which holds for small $\left|x\right|$, it is possible to infer the
decay law of the BG beams intensity in $\rho=\bar{\rho}$ at large
$z$: $I_{BG}\left(z\right)\propto z^{-2\left|m\right|-2}$.

\begin{figure}
\noindent \begin{centering}
\includegraphics[width=1\textwidth]{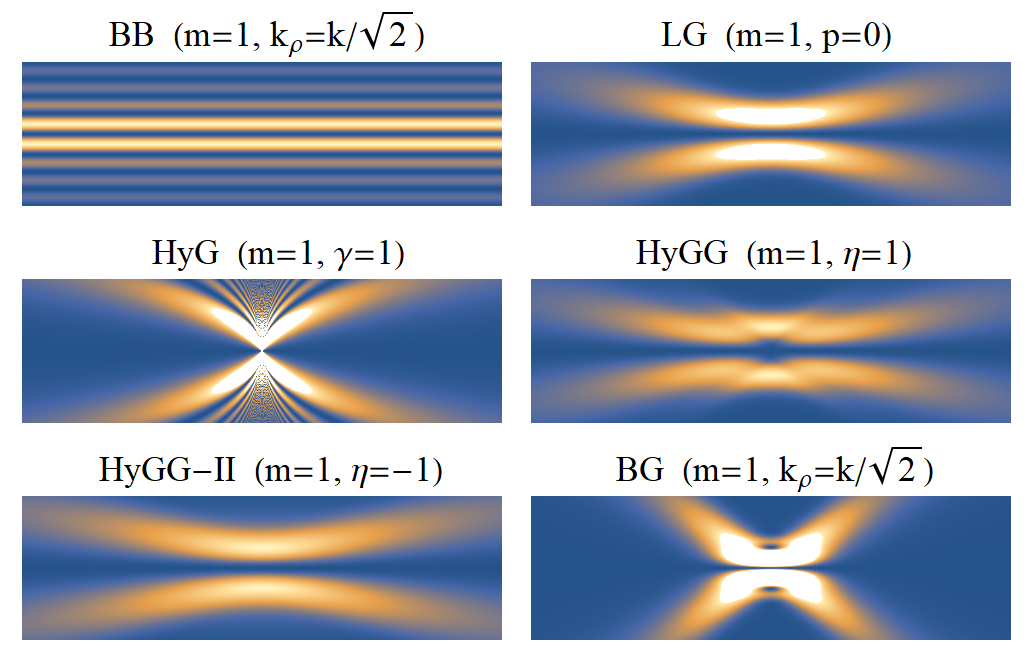}
\par\end{centering}
\caption{Comparison of the longitudinal intensity profiles ($x$-coordinate versus $z$-coordinate) of some paraxial beams for $w_0=\lambda$.\label{fig:longProfiles}}
\end{figure}

To conclude the section, a comparison of the longitudinal intensity 
profiles of some representative solutions belonging to each of the 
above described paraxial families is reported in Figure \ref{fig:longProfiles}.
As it can be noticed, the energy distribution of these scalar waves
as a function of the propagation coordinate varies considerably 
from beam to beam. 

\section{Vortex solutions of the vector wave equation\label{sec:Vector}}

As shown in the two previous sections, exact and paraxial solutions
of the scalar wave equation which are eigenfunctions of the $\hat{J}_{z}$
operator can be easily derived by means of standard mathematical techniques;
indeed, such scalar OAM solutions have been deeply explored in the
literature. However, an electromagnetic field can hardly be described
in terms of a pure scalar function of space and time, since only a
full vector treatment allows to take into account the presence of
complex polarization structures. Complete solutions of the vector
wave equation that can be directly applied to boundary-value problems
in electromagnetism have been derived for certain separable systems
of cylindrical coordinates and for spherical coordinates: in this
context, the electromagnetic field can be resolved into two partial
fields, each derivable from a function satisfying the scalar wave
equation \cite{Morse1953,Stratton1941}. 

Within any closed domain of a homogeneous and isotropic medium with
zero conductivity and free-charge density, all vectors characterizing
the electromagnetic field satisfy:
\begin{equation}
\nabla^{2}\boldsymbol{\Psi}-\epsilon\mu\frac{\partial^{2}\boldsymbol{\Psi}}{\partial t^{2}}=0,\label{eq:d'Alembert}
\end{equation}
being $\epsilon$ and $\mu$ the inductive capacities of the medium.
On using the following relation:
\begin{equation}
\nabla^{2}\boldsymbol{\Psi}=\nabla\left(\nabla\cdot\boldsymbol{\Psi}\right)-\nabla\times\nabla\times\boldsymbol{\Psi}
\end{equation}
and assuming a time-harmonic dependence of the form $\exp\left(i\omega t\right)$,
equation (\ref{eq:d'Alembert}) becomes:
\begin{equation}
\nabla\left(\nabla\cdot\boldsymbol{\psi}\right)-\nabla\times\nabla\times\boldsymbol{\psi}+k^{2}\boldsymbol{\psi}=0,\label{eq:vectorHelmholtzEquation}
\end{equation}
with $k=\epsilon\mu\omega^{2}$. Equation (\ref{eq:vectorHelmholtzEquation})
represents the vector analog of (\ref{eq:HelmholtzEquation}). As
emphasized by Julius A. Stratton in \cite{Stratton1941}, equation
(\ref{eq:vectorHelmholtzEquation}) can always be replaced by a system
of three scalar equations, but it is only when $\boldsymbol{\psi}$
is expressed in rectangular components that three independent scalar
equations are obtained: 
\begin{equation}
\nabla^{2}\psi_{x}+k^{2}\psi_{x}=0;\quad\nabla^{2}\psi_{y}+k^{2}\psi_{y}=0;\quad\nabla^{2}\psi_{z}+k^{2}\psi_{z}=0.\label{eq:3DscalarEquation}
\end{equation}
The Laplacian in (\ref{eq:3DscalarEquation}) can of course be written
in different coordinate systems, but the vector character of the original
equation is inevitably lost. Three independent vector solutions of
(\ref{eq:vectorHelmholtzEquation}) can instead be built as follows:
\begin{equation}
\mathbf{M}=\nabla\times\mathbf{C}g;\quad\mathbf{N}=\frac{1}{k}\nabla\times\mathbf{M};\quad\mathbf{L}=\nabla g,\label{eq:vectorFunctions}
\end{equation}
being $g=g\left(x_{1},x_{2},x_{3}\right)$ a scalar function satisfying
(\ref{eq:HelmholtzEquation}) and $\mathbf{C}$ a unit norm constant
vector. The three vector functions above possess several interesting
properties: $\mathbf{M}=\mathbf{L}\times\mathbf{C}=k^{-1}\nabla\times\mathbf{N}$,
$\mathbf{L}\cdot\mathbf{M}=0$, $\nabla\times\mathbf{L}=0$,
$\nabla\cdot\mathbf{L}=-k^{2}g$; moreover, it is easy to prove
that both $\mathbf{M}$ and $\mathbf{N}$ are solenoidal and,
owing to this and to the fact of being each proportional to the curl
of the other, can be employed to represent the electric and magnetic
fields. 

Since a decomposition of any arbitrary vector wavefunction in terms
of (\ref{eq:vectorFunctions}) is completely general, families of
orthogonal vector solutions of the wave equation (\ref{eq:d'Alembert})
are derived via (\ref{eq:vectorFunctions}) from sets of solutions
of the scalar wave equation (\ref{eq:waveEquation}) in the various
coordinate systems. Also the field polarization is now being considered
and care must be taken to properly describe the action of the rotation
group generators on the whole vector field. 

As can be shown by expanding to first order an infinitesimal rotation
about the $z$-axis (see \ref{sec:AMoperator}), the complete form
of the $\hat{J}_{z}$ operator for vector fields is provided by:
\begin{equation}
i\hat{J}_{z}=-i\frac{\partial}{\partial\varphi}+iS_{z},\label{eq:Jz}
\end{equation}
where $S_{z}$ corresponds to the third SO(3) generator in three-dimensional
matrix representation. When the vector field is resolved into its
Cartesian components, the explicit expression for $S_{z}$ is given
by $J_{3}$ in (\ref{eq:Jk}) and equation (\ref{eq:Jz}) reduces
to its more common form \cite{Rose1957}:
\begin{equation}
i\hat{J}_{z}=-i\left(x\frac{\partial}{\partial y}-y\frac{\partial}{\partial x}\right)+\left[\begin{array}{ccc}
0 & -i & 0\\
i & 0 & 0\\
0 & 0 & 0
\end{array}\right].\label{eq:JzCartesian}
\end{equation}

Let $\left\{ g_{m}\right\} =\left\{ \varPhi_{m}\left(\varphi\right)u\left(x_{1},x_{3}\right)\right\} $
be a set of OAM scalar functions derived according to Section \ref{sec:Scalar}.
A general vector solution of (\ref{eq:vectorHelmholtzEquation}) can
be constructed from $\left\{ g_{m}\right\} $ via the superposition:
\begin{equation}
\boldsymbol{\psi}=\frac{i}{\omega}\sum_{m}\left(a_{m}\mathbf{M}_{m}+b_{m}\mathbf{N}_{m}+c_{m}\mathbf{L}_{m}\right),\label{eq:vectorExpansion}
\end{equation}
where $a_{m}$, $b_{m}$ and $c_{m}$ represent suitable expansion
coefficients and the vector fields in the set $\left\{ \mathbf{M}_{m},\mathbf{N}_{m},\mathbf{L}_{m}\right\} $
are provided by (\ref{eq:vectorFunctions}) with $g=g_{m}$. Therefore,
the transformation of $\mathbf{\boldsymbol{\psi}}$ under an infinitesimal
rotation about the $z$-axis leads back to the action of the $\hat{J}_{z}$
operator on each basis function. In all the five separable coordinate
systems for which the $\varphi$-coordinate exists, the following
relations hold:
\begin{equation}
i\hat{J}_{z}\mathbf{M}_{m}=m\mathbf{M}_{m};\quad i\hat{J}_{z}\mathbf{N}_{m}=m\mathbf{N}_{m};\quad i\hat{J}_{z}\mathbf{L}_{m}=m\mathbf{L}_{m},\label{eq:vectorEigenfunctions}
\end{equation}
as explicitly shown in \ref{sec:AMoperator}. The importance of this
last result lies in the fact that it is the only requirement of separability
of the scalar Helmholtz equation which ensures the vector fields $\left\{ \mathbf{M}_{m},\mathbf{N}_{m},\mathbf{L}_{m}\right\} $
built from the OAM set $\left\{ g_{m}\right\} $ to be eigenfunctions
of the $\hat{J}_{z}$ operator.

From (\ref{eq:vectorFunctions}), (\ref{eq:Jz}) and (\ref{eq:vectorEigenfunctions})
we understand that the so-derived eigenfunctions cannot be simply
considered as vector OAM waves, since their structure is in general
more complex, also involving polarization and thus a SAM contribution.
Furthermore, for all the three eigenfunctions in (\ref{eq:vectorEigenfunctions}),
it is the vector field components which present the characteristic
vortex term $\varPhi_{m}\left(\varphi\right)$, meaning that the phase
evolution around the $z$-axis is accompanied by the evolution of
the state of polarization. Such eigenfunctions are then named \textit{vector
vortex beams} (VVBs\nomenclature{VVB}{vector vortex beam}), bearing
in mind that in this case the on-axis phase singularity is no longer
exclusive and also polarization singularities may be present (see,
for instance, \cite{Berry2001,Dennis2002,Freund2002,Nye1999,Wolf2009}
and references therein). The above described approach undoubtedly
provides a rigorous method for understanding how the VVBs definition
can be traced back to general symmetry arguments.

In order to give some instructive examples, let us suppose that the
vector potential $\mathbf{A}$ can be represented by an expansion
in characteristic vector functions, as in (\ref{eq:vectorExpansion}).
The electric field $\mathbf{E}$ in the Lorentz gauge is then
expressed by:
\begin{equation}
\mathbf{E}=-\sum_{m}\left(a_{m}\mathbf{M}_{m}+b_{m}\mathbf{N}_{m}\right),\label{eq:electricFieldExpansion}
\end{equation}
the usual time-harmonic term $\exp\left(i\omega t\right)$ being implied.
Considering a circular cylindrical coordinate system, for which the
natural OAM scalar field basis $\left\{ g_{m}\right\} $ is provided
by BBs, and choosing the $z$-directed unit vector as $\mathbf{C}$
in (\ref{eq:vectorFunctions}), we get:
\begin{equation}
\mathbf{M}_{m}=\frac{k_{\rho}}{2}\left[i\left(J_{m-1}+J_{m+1}\right)\mathbf{u}_{\rho}-\left(J_{m-1}-J_{m+1}\right)\mathbf{u}_{\varphi}\right]\mathrm{e}^{-ik_{z}z+im\varphi};\label{eq:Mm}
\end{equation}
\begin{eqnarray}
\mathbf{N}_{m} & = & \frac{k_{\rho}}{2k}\left[-ik_{z}\left(J_{m-1}-J_{m+1}\right)\mathbf{u}_{\rho}+k_{z}\left(J_{m-1}+J_{m+1}\right)\mathbf{u}_{\varphi}\right.\nonumber \\
 & + & \left.2k_{\rho}J_{m}\mathbf{u}_{z}\right]\exp\left(-ik_{z}z+im\varphi\right),\label{eq:Nm}
\end{eqnarray}
where all Bessel functions have argument $\left(k_{\rho}\rho\right)$,
$\left\{ \mathbf{u}_{\rho},\mathbf{u}_{\varphi},\mathbf{u}_{z}\right\} $
represent the cylindrical system unit vectors and (\ref{eq:BesselModes})
has been employed together with some known properties of Bessel functions
under the assumption $m\geq0$, for convenience. 

\begin{figure}[t]
\noindent \begin{centering}
\includegraphics[width=1\textwidth]{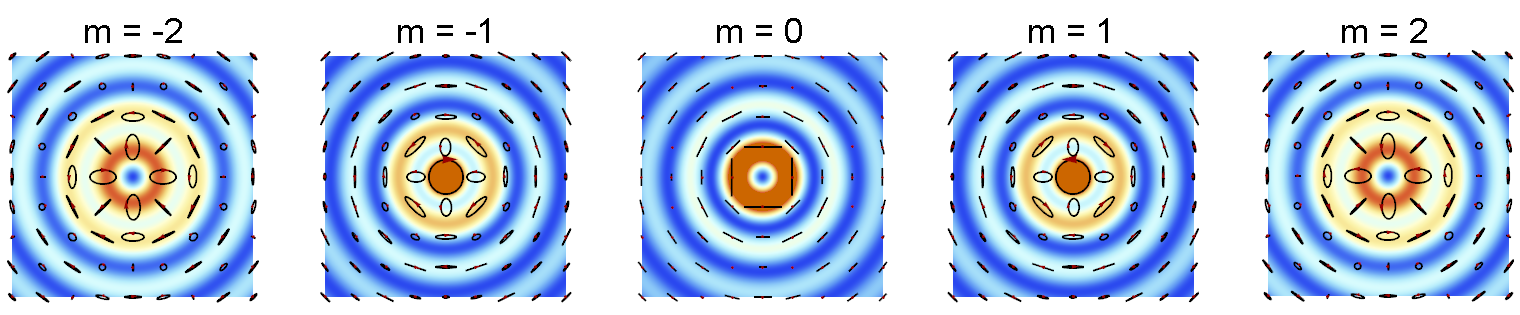}
\par\end{centering}
\caption{Norm and polarization plots ($y$-coordinate versus $x$-coordinate) of the function $\mathbf{M}_{m}$ defined
in (\ref{eq:Mm}) for $k_{\rho}=k/\sqrt{2}$.\label{fig:vectorProfiles}}
\end{figure}

In Figure \ref{fig:vectorProfiles}, the norm and polarization plots
of (\ref{eq:Mm}) are reported for some values of $m$, revealing
two apparent deviations from the conventional OAM patterns: when $m=0$,
a null is present instead of the usual on-axis intensity maximum,
whereas for $m=1$ the central null is replaced by a maximum.
Both deviations are attributable to the mixing of spatial and polarization
contributions, which also affects all the other modes, as evidenced
by the circumstance that additional factors $\mathrm{e}^{\pm i\varphi}$
appear once (\ref{eq:Mm}) is rewritten in terms of the Cartesian
unit vectors. It should be noted that the presence of the on-axis
intensity null in the $m=0$ case results from a polarization singularity
known as \textit{V point} in singular optics \cite{Freund2002}. 

In \cite{VolkeSepulveda2002}, Karen Volke-Sep\'{u}lveda et al. have shown
that equations (\ref{eq:electricFieldExpansion}), (\ref{eq:Mm})
and (\ref{eq:Nm}) make it possible to describe a wide variety of
vector solutions in different polarization states. For instance, a
radially or azimuthally polarized beam can be obtained by setting
$a_{m}=0$ and $b_{m}=i\delta_{m0}E_{0}k/\left(k_{\rho}k_{z}\right)$
or $a_{m}=-\delta_{m0}E_{0}/k_{\rho}$ and $b_{m}=0$ in (\ref{eq:electricFieldExpansion}),
respectively; here $E_{0}$ represents a constant proportional to
the square root of the beam power and with electric field units. Such
peculiar vector waves with circular symmetric structure can be seen
as free-space transverse magnetic (TM\nomenclature{TM}{transverse magnetic})
and transverse electric (TE\nomenclature{TE}{transverse electric})
modes, as can be deduced from their electric field expressions:
\begin{equation}
\mathbf{E}_{R}=E_{0}\left[J_{1}\left(k_{\rho}\rho\right)\mathbf{u}_{\rho}-i\left(\frac{k_{\rho}}{k_{z}}\right)J_{0}\left(k_{\rho}\rho\right)\mathbf{u}_{z}\right]\exp\left(-ik_{z}z+i\omega t\right);\label{eq:radialPolarizedBeam}
\end{equation}
\begin{equation}
\mathbf{E}_{A}=E_{0}\left[J_{1}\left(k_{\rho}\rho\right)\mathbf{u}_{\varphi}\right]\exp\left(-ik_{z}z+i\omega t\right),\label{eq:azimuthalPolarizedBeam}
\end{equation}
where the subscripts $R$ and $A$ stand for radial and azimuthal,
respectively, and the time dependence is now explicitly written. Both
(\ref{eq:radialPolarizedBeam}) and (\ref{eq:azimuthalPolarizedBeam})
are eigenfunctions of the $\hat{J}_{z}$ operator with eigenvalue
$m=0$, by construction. 

\begin{figure}[t]
\noindent \begin{centering}
\includegraphics[width=1\textwidth]{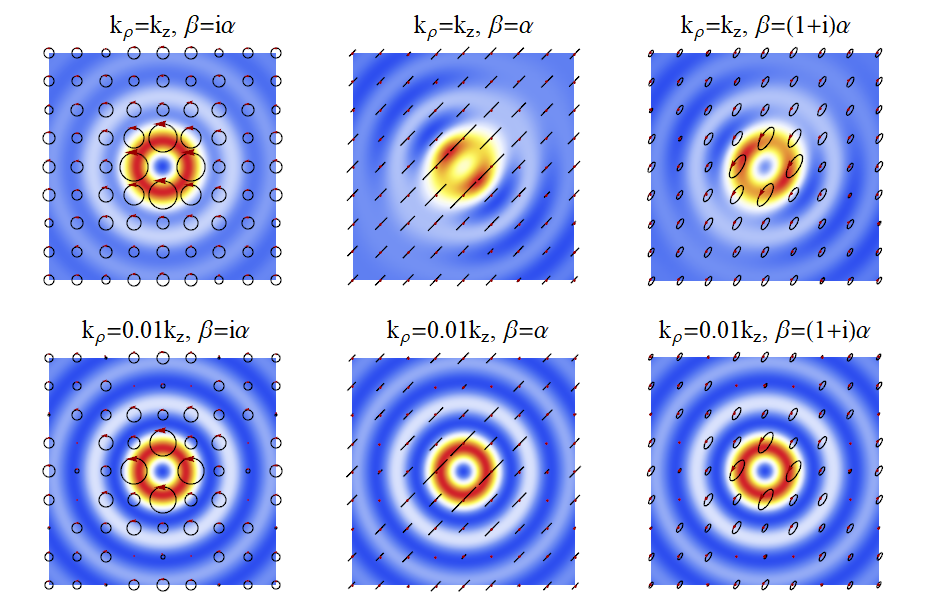}
\par\end{centering}
\caption{Norm and polarization plots ($y$-coordinate versus $x$-coordinate) of the electric field $\mathbf{E}$ in (\ref{eq:genPolState}) for some nonparaxial (upper row) and paraxial 
(lower row) solutions with $m=1$.\label{fig:genPolState}}
\end{figure}

Following the approach presented in \cite{VolkeSepulveda2002}, right-handed
and left-handed circular polarized modes are easily derived from $b_{m}=\mp a_{m}k/k_{z}$
upon proper choice of the $a_{m}$ coefficients:
\begin{equation}
\mathbf{E}_{\pm}=E_{0}\left[J_{m}\left(k_{\rho}\rho\right)\left(i\mathbf{u}_{x}\mp\mathbf{u}_{y}\right)\mp\left(\frac{k_{\rho}}{k_{z}}\right)\mathrm{e}^{\pm i\varphi}J_{m\pm1}\left(k_{\rho}\rho\right)\mathbf{u}_{z}\right]\mathrm{e}^{-ik_{z}z+im\varphi+i\omega t},\label{eq:circularPolarizedBeam}
\end{equation}
being $\left\{ \mathbf{u}_{x},\mathbf{u}_{y},\mathbf{u}_{z}\right\} $
the Cartesian unit vectors. In particular, since equation (\ref{eq:circularPolarizedBeam})
is obtained from the $\left(m\pm1\right)$ term of (\ref{eq:electricFieldExpansion}),
we find:
\begin{equation}
i\hat{J}_{z}\mathbf{E}_{\pm}=\left(m\pm1\right)\mathbf{E}_{\pm},
\end{equation}
as can be inferred from (\ref{eq:vectorEigenfunctions}) or explicitly
verified using (\ref{eq:JzCartesian}). Whereas the two orthogonal
$x$-polarized and $y$-polarized states are simple superpositions
of the right-circular and left-circular ones, the most general field
is given in terms of the Jones vector Cartesian components $\alpha$
and $\beta$ \cite{Simmons1970}:
\begin{eqnarray}
\mathbf{E} & = & E_{0}\left\{ \left(\frac{-ik_{\rho}}{2k_{z}}\right)\left[\left(\alpha+i\beta\right)\mathrm{e}^{-i\varphi}J_{m-1}\left(k_{\rho}\rho\right)-\left(\alpha-i\beta\right)\mathrm{e}^{i\varphi}J_{m+1}\left(k_{\rho}\rho\right)\right]\mathbf{u}_{z}\right.\nonumber \\
 & + & \left.\left[\alpha\mathbf{u}_{x}+\beta\mathbf{u}_{y}\right]J_{m}\left(k_{\rho}\rho\right)\frac{}{}\right\} \exp\left(-ik_{z}z+im\varphi+i\omega t\right).\label{eq:genPolState}
\end{eqnarray}
For $\beta=\pm i\alpha$, equation (\ref{eq:genPolState}) leads back to 
(\ref{eq:circularPolarizedBeam}). Any other choice of the polarization 
parameters $\alpha$ and $\beta$ in (\ref{eq:genPolState}) does not give 
rise to an eigenvector of the $\hat{J}_{z}$ operator, as evidenced by the
fact that the circular symmetry of both the beam energy and polarization 
is broken (see Figure \ref{fig:genPolState}). The angular momentum density for the various cylindrical 
solutions described above has been rigorously calculated in \cite{VolkeSepulveda2002},
showing that its distribution is indeed not radially symmetric in the case
of linearly polarized BBs. On the other hand, the presence of the ratio $\left(k_{\rho}/k_{z}\right)$
in all previous formulas makes it possible to extend the analysis
to the paraxial regime $k_{\rho}\ll k_{z}$, with which most
of the literature is concerned. Under this approximation, the electric field $\mathbf{E}$ 
in (\ref{eq:genPolState}) is found to be eigenvector of $-i\partial/\partial\varphi$,
the orbital component of $i\hat{J}_{z}$, and the circular symmetry of the solution norm is restored, 
as it is shown in Figure \ref{fig:genPolState}.
Despite the analytic expressions for paraxial VVBs may be obtained more practically 
by non-separable combinations of spatial and polarization modes \cite{Andrews2013}, 
the above derivation could be in some circumstances preferable. 

Finally, among the various paraxial and non-paraxial VVBs, vector
beams presenting cylindrical symmetric states of polarization, often
known as \textit{cylindrical vector} (CV\nomenclature{CV}{cylindrical vector})
\textit{beams} or vortices, have been extensively analyzed due to their interesting
properties such as the tighter waist upon focusing (a thorough review
of the applications can be found in \cite{Zhan2009}).

\section{Conclusions}

In this work, an in-depth overview on the exact and paraxial vortex
solutions of the homogeneous wave equation has been presented. 
These solutions have been interpreted as eigenstates of the 
$\hat{J}_{z}$ operator, leading to a mixture of polarization
and spatial modes in the case of a general vector vortex wave.

Whereas a scalar approach is often convenient to describe the transverse
free-space modes of linearly polarized laser radiation in optics, the same
does not hold for the radio frequency domain, where a full vector treatment
is usually mandatory. In neither cases could the homogeneous wave equation
probably be sufficient for the most realistic modeling of the OAM waveforms
generated experimentally by means of resonators and antennas, for which
only the complete Maxwell equations with real or equivalent source terms
can ideally provide the highest level of accuracy. Nevertheless, it is unquestionably
true that the homogeneous Helmholtz equation still represents the
most affordable way for describing specific free-space solutions as well as for
providing general basis sets to expand any arbitrary waveform. For instance,
the scalar and vector OAM solutions analyzed in this paper could
be easily interpreted as free-space generalizations of resonator and waveguide
eigenmodes that are usually found in both laser theory and electromagnetics
handbooks. The intensity, phase, polarization and propagation properties
of such free-space beams are therefore considered as ideal reference models
or, alternatively, as targets to be reproduced experimentally up to a certain
degree of approximation via proper synthesis methods.
Since many of the exact vortex solutions possess infinite transverse extension, 
are divergence-free and carry infinite energy,
paraxial OAM waves with Gaussian envelope or truncated forms which take
into account diffraction and finite-size effects must be preferred in order to
ensure a more realistic description.

\ack
The main ideas and results of this work were carried out during my
PhD in physics at the University of Torino. Among the various people to whom I am 
indebted, I would like to especially thank Prof. Roberto Tateo, Prof. Paolo Gambino 
and Prof. Miguel Onorato, from the Department of Physics, Dr. Rossella Gaffoglio, 
Prof. Francesco Andriulli and Prof. Giuseppe Vecchi, from the Polytechnic University
of Torino, Dr. Assunta De Vita and Eng. Bruno Sacco, from the Centre for Research 
and Technological Innovation, RAI Radiotelevisione Italiana.

\appendix

\section{Vortex solutions in different coordinate systems\label{sec:A1}}

\subsection{Spherical coordinates}

When expressed in a spherical coordinate system, the scalar Helmholtz
equation (\ref{eq:HelmholtzEquation}) reads: 
\begin{equation}
\frac{1}{r}\frac{\partial^{2}\left(r\psi\right)}{\partial r^{2}}+\frac{1}{r^{2}\sin\theta}\frac{\partial}{\partial\theta}\left(\sin\theta\frac{\partial\psi}{\partial\theta}\right)+\frac{1}{r^{2}\sin^{2}\theta}\frac{\partial^{2}\psi}{\partial\phi^{2}}+k^{2}\psi=0.\label{eq:HelmholtzSferica}
\end{equation}
This section is devoted to the search for spherical OAM solutions
of the form:
\begin{equation}
\psi\left(r,\theta,\phi\right)=C\,\frac{\exp\left(-ikr+im\phi\right)}{r}R\left(r\right)\Theta\left(\theta\right),\label{eq:OndaSfericaOAM}
\end{equation}
where $R\left(r\right)$ and $\Theta\left(\theta\right)$ represent
functions to be determined and $C$ is a constant. It is immediate to verify that, in order
for (\ref{eq:OndaSfericaOAM}) to satisfy equation (\ref{eq:HelmholtzSferica}),
$\Theta$ cannot be constant. As a first step, for simplicity, it
can be assumed that $R=1$, so that equation (\ref{eq:HelmholtzSferica})
reduces to:
\begin{equation}
\frac{d^{2}\Theta\left(\theta\right)}{d\theta^{2}}+\cot\theta\frac{d\Theta\left(\theta\right)}{d\theta}-m^{2}\left(\csc\theta\right)^{2}\Theta\left(\theta\right)=0,
\end{equation}
provided that $r\neq0$. By introducing the parametric coordinate
$t=\tan\left(\theta/2\right)$, from which $\cos\theta=\frac{1-t^{2}}{1+t^{2}}$
and $\sin\theta=\frac{2t}{1+t^{2}}$ follow, we get:
\begin{equation}
\frac{d^{2}\Theta\left(t\right)}{dt^{2}}+\frac{1}{t}\frac{d\Theta\left(t\right)}{dt}-\frac{m^{2}}{t^{2}}\Theta\left(t\right)=0,\label{eq:EquazionePolare}
\end{equation}
a Fuchsian ordinary differential equation with singularity in $t=0$.
Since the roots of the corresponding indicial equation are $\pm\left|m\right|$,
the two linearly independent solutions of (\ref{eq:EquazionePolare})
can be written in the following way:
\begin{equation}
\begin{cases}
\Theta^{(1)}\left(t\right)=\sum_{k=0}^{\infty}c_{k}\,t^{k+\left|m\right|};\\
\Theta^{(2)}\left(t\right)=\sum_{k=0}^{\infty}d_{k}\,t^{k-\left|m\right|}+d\,\Theta^{(1)}\left(t\right)\log\left(t\right).
\end{cases}
\end{equation}
From substitution of the power series in (\ref{eq:EquazionePolare})
we easily find $k,\,d=0$, hence $\Theta^{(1)}\left(t\right)=t^{\left|m\right|}$
and $\Theta^{(2)}\left(t\right)=t^{-\left|m\right|}$, where only
the first represents a regular solution in $t=0$. Then, expression
(\ref{eq:OndaSfericaOAM}) reduces to:
\begin{equation}
\psi\left(r,\theta,\phi\right)=C\,\frac{\exp\left(-ikr+im\phi\right)}{r}\left[\tan\left(\frac{\theta}{2}\right)\right]^{\left|m\right|},\label{eq:HelicoPolarBeam}
\end{equation}
which is not defined for $\theta=\pi$. A regularized
version of the previous function over the whole sphere can be provided
through the use of both $\Theta^{(1)}$ and $\Theta^{(2)}$, for instance:
\begin{equation}
\psi_{reg}\left(r,\theta,\phi\right)=C\,\frac{\exp\left(-ikr+im\phi\right)}{r}\begin{cases}
\left[\tan\left(\frac{\theta}{2}\right)\right]^{\left|m\right|} & \mathrm{for}\,\theta\in\left[0,\frac{\pi}{2}\right];\\
\left[\tan\left(\frac{\theta}{2}\right)\right]^{-\left|m\right|} & \mathrm{for}\,\theta\in\left(\frac{\pi}{2},\pi\right].
\end{cases}\label{eq:HelicoPolarBeamRegolarizzato}
\end{equation}

OAM waves of the form (\ref{eq:HelicoPolarBeam}) and (\ref{eq:HelicoPolarBeamRegolarizzato})
may be called ``helico-polar'' modes, as they possess helical wavefronts
and show a very simple dependence on the polar angle $\theta$. It
can be shown that these non-paraxial waves represent square-integrable
solutions with respect to the spherical measure $\sin\theta d\theta d\phi$.

In order to prove that the well-known spherical multipoles correspond
to generalizations of the above derived waves, the radial term $R\left(r\right)$
in (\ref{eq:OndaSfericaOAM}) is now reintroduced. Equation (\ref{eq:HelmholtzSferica})
becomes:
\begin{equation}
-\frac{2ik}{r}\dot{R}\Theta+\frac{1}{r}\ddot{R}\Theta+\frac{\cot\theta}{r^{3}}R\dot{\Theta}+\frac{1}{r^{3}}R\ddot{\Theta}{\displaystyle -\frac{m^{2}\left(\csc\theta\right)^{2}}{r^{3}}R\Theta=0},
\end{equation}
where the simplified notation $\dot{X}\left(x\right)=dX\left(x\right)/dx$
has been employed. Then, for $r\neq0$ we find:
\begin{equation}
\begin{cases}
\ddot{R}-2ik\dot{R}+\frac{L}{r^{2}}R=0;\\
\ddot{\Theta}+\cot\theta\dot{\Theta}-\left[m^{2}\left(\csc\theta\right)^{2}+L\right]\Theta=0,
\end{cases}\label{eq:Sistema}
\end{equation}
being $L$ a constant term. To solve the first equation we set $R=\sqrt{r}\mathcal{R}\left(r\right)\mathrm{e}^{ikr}$,
which leads to Bessel equation in $kr$:
\begin{equation}
r^{2}\mathcal{\ddot{R}}+r\mathcal{\dot{R}}+\left(k^{2}r^{2}+L-\frac{1}{4}\right)\mathcal{R}=0,
\end{equation}
and therefore:
\begin{equation}
\begin{cases}
R^{(1)}=\exp\left(ikr\right)\sqrt{r}\,J_{\frac{1}{2}\sqrt{1-4L}}\left(kr\right);\\
R^{(2)}=\exp\left(ikr\right)\sqrt{r}\,Y_{\frac{1}{2}\sqrt{1-4L}}\left(kr\right)
\end{cases}
\end{equation}
represent the two linearly independent radial solutions, given in
terms of Bessel functions of the first and second kind, respectively.
After performing the substitution $\cos\theta=t$, the second equation
in (\ref{eq:Sistema}) is easily reduced to:
\begin{equation}
\left(1-t^{2}\right)\frac{d^{2}\Theta\left(t\right)}{dt^{2}}-2t\frac{d\Theta\left(t\right)}{dt}-\left(L+\frac{m^{2}}{1-t^{2}}\right)\Theta\left(t\right)=0,
\end{equation}
which admits solutions in terms of the generalized Legendre polynomials.
Finally, by imposing the regularity of $\Theta$ at the boundary points,
it can be proven that $L=-\ell\left(\ell+1\right)$ for an integer
$\ell\geq\left|m\right|$ and equation (\ref{eq:OndaSfericaOAM})
acquires the standard form:
\begin{equation}
\psi\left(r,\theta,\phi\right)=b_{\ell}\left(kr\right)Y_{\ell}^{m}\left(\theta,\phi\right),
\end{equation}
where $b_{\ell}\left(kr\right)$ corresponds to a linear combination
of the two independent spherical Bessel functions:
\begin{equation}
j_{\ell}\left(kr\right)=\sqrt{\frac{\pi}{2kr}}J_{\ell+1/2}\left(kr\right);\quad y_{\ell}\left(kr\right)=\sqrt{\frac{\pi}{2kr}}Y_{\ell+1/2}\left(kr\right)
\end{equation}
and $Y_{\ell}^{m}$ are the spherical harmonics.

\subsection{Prolate spheroidal coordinates}

Let $u\left(\xi,\eta\right)=A\left(\xi\right)B\left(\eta\right)$
be a function that satisfies (\ref{eq:reducedEquation}) in the prolate
spheroidal coordinate system $\left\{ \xi,\varphi,\eta\right\} $.
On using the explicit expressions of the metric scale factors \cite{Morse1953}:
\begin{equation}
\begin{cases}
h_{1}=h_{\xi}=\varsigma\,\sqrt{\sinh^{2}\xi+\sin^{2}\eta};\\
h_{2}=h_{\varphi}=\varsigma\,\sinh\xi\sin\eta;\\
h_{3}=h_{\eta}=\varsigma\,\sqrt{\sinh^{2}\xi+\sin^{2}\eta},
\end{cases}
\end{equation}
where $2\varsigma$ is the distance between the foci of the family
of confocal ellipses and hyperbolas, we obtain two forms of spheroidal
wave equations:
\begin{equation}
\begin{cases}
\ddot{A}+\dot{A}\coth\xi+\left(-\tau+\varsigma^{2}k^{2}\sinh^{2}\xi-m^{2}/\sinh^{2}\xi\right)A=0;\\
\ddot{B}+\dot{B}\cot\eta+\left(\tau+\varsigma^{2}k^{2}\sin^{2}\eta-m^{2}/\sin^{2}\eta\right)B=0,
\end{cases}
\end{equation}
with solutions given by \cite{Miller1977}:
\begin{equation}
\begin{cases}
A\left(\xi\right)=S_{n}^{\left|m\right|}\left(\cosh\xi,\varsigma^{2}k^{2}\right);\\
B\left(\eta\right)=S_{n}^{\left|m\right|}\left(\cos\eta,\varsigma^{2}k^{2}\right),
\end{cases}
\end{equation}
where $S_{n}^{\left|m\right|}$ represents a spheroidal wavefunction,
$n\geq\left|m\right|$ is an integer and the discrete eigenvalues
$\tau=\tau_{n}^{\left|m\right|}$ are analytic functions of $\varsigma^{2}k^{2}$.

\subsection{Oblate spheroidal coordinates}

In the oblate spheroidal coordinate system $\left\{ \xi,\varphi,\eta\right\} $,
the metric scale factors read \cite{Morse1953}: 
\begin{equation}
\begin{cases}
h_{1}=h_{\xi}=\varsigma\,\sqrt{\sinh^{2}\xi+\sin^{2}\eta};\\
h_{2}=h_{\varphi}=\varsigma\,\cosh\xi\cos\eta;\\
h_{3}=h_{\eta}=\varsigma\,\sqrt{\sinh^{2}\xi+\sin^{2}\eta}
\end{cases}
\end{equation}
and equation (\ref{eq:reducedEquation}) can be separated into:
\begin{equation}
\begin{cases}
\ddot{A}+\dot{A}\tanh\xi+\left(-\tau+\varsigma^{2}k^{2}\cosh^{2}\xi+m^{2}/\cosh^{2}\xi\right)A=0;\\
\ddot{B}+\dot{B}\cot\eta+\left(\tau-\varsigma^{2}k^{2}\sin^{2}\eta-m^{2}/\sin^{2}\eta\right)B=0,
\end{cases}
\end{equation}
whose bounded and single-valued solutions are written as \cite{Miller1977}:
\begin{equation}
\begin{cases}
A\left(\xi\right)=S_{n}^{\left|m\right|}\left(-i\sinh\xi,\varsigma^{2}k^{2}\right);\\
B\left(\eta\right)=S_{n}^{\left|m\right|}\left(\cos\eta,-\varsigma^{2}k^{2}\right).
\end{cases}
\end{equation}

\subsection{Parabolic coordinates}

In parabolic coordinates $\left\{ \mu,\varphi,\nu\right\} $, with
metric scale factors \cite{Morse1953}: 
\begin{equation}
\begin{cases}
h_{1}=h_{\mu}=\sqrt{\mu^{2}+\nu^{2}};\\
h_{2}=h_{\varphi}=\mu\nu;\\
h_{3}=h_{\nu}=\sqrt{\mu^{2}+\nu^{2}},
\end{cases}
\end{equation}
$u\left(\mu,\nu\right)=U\left(\mu\right)V\left(\nu\right)$ and equation
(\ref{eq:reducedEquation}) reduces to:
\begin{equation}
\begin{cases}
\ddot{U}+\dot{U}\mu^{-1}+\left(k^{2}\mu^{2}-m^{2}/\mu^{2}-\tau\right)U=0;\\
\ddot{V}+\dot{V}\nu^{-1}+\left(k^{2}\nu^{2}-m^{2}/\nu^{2}+\tau\right)V=0.
\end{cases}
\end{equation}
The separated solutions take the form \cite{Miller1977}:
\begin{equation}
\begin{cases}
U\left(\mu\right)=\mu^{m}\exp\left(\pm\frac{ik\mu^{2}}{2}\right)\,_{1}F_{1}\left(\frac{i\tau}{4k}+\frac{m+1}{2},m+1;\mp ik\mu^{2}\right);\\
V\left(\nu\right)=\nu^{m}\exp\left(\pm\frac{ik\nu^{2}}{2}\right)\,_{1}F_{1}\left(-\frac{i\tau}{4k}+\frac{m+1}{2},m+1;\mp ik\nu^{2}\right),
\end{cases}
\end{equation}
where $_{1}F_{1}$ is the Kummer hypergeometric confluent function. 

\section{Angular momentum operator\label{sec:AMoperator}}

Let $\boldsymbol{\psi}\left(x_{1},\varphi,x_{3}\right)$ be an arbitrary
vector function in one of the five coordinate systems $\left\{ x_{1},x_{2}=\varphi,x_{3}\right\} $
for which equation (\ref{eq:reducedEquation}) separates. Under infinitesimal
rotation by an angle $\varepsilon$ about the $z$-axis, the function
$\boldsymbol{\psi}$ transforms as:
\begin{equation}
\boldsymbol{\psi}'\left(x_{1}',\varphi',x_{3}'\right)=\boldsymbol{\psi}\left(x_{1},\varphi-\varepsilon,x_{3}\right)\sim\sum_{i=1}^{3}\left[\psi_{i}\left(x_{1},\varphi,x_{3}\right)-\varepsilon\frac{\partial\psi_{i}\left(x_{1},\varphi,x_{3}\right)}{\partial\varphi}\right]\mathbf{u}_{i},\label{eq:infinitesimalTransf}
\end{equation}
where $\left\{ \mathbf{u}_{1},\mathbf{u}_{2}=\mathbf{u}_{\varphi},\mathbf{u}_{3}\right\} $
correspond to the standard unit vectors of the considered system and
a first order Taylor expansion in $\varepsilon$ has been performed.
Equation (\ref{eq:infinitesimalTransf}) can be rewritten in the following
form: 
\begin{equation}
\boldsymbol{\psi}'\left(x_{1}',\varphi',x_{3}'\right)\sim\left[1-\varepsilon\frac{\partial}{\partial\varphi}\right]\left(\sum_{i=1}^{3}\psi_{i}\mathbf{u}_{i}\right)+\varepsilon\sum_{i=1}^{3}\psi_{i}\frac{\partial\mathbf{u}_{i}}{\partial\varphi}.\label{eq:psiPrime}
\end{equation}
On using some basic results of differential geometry, it can be shown
that \cite{Morse1953}: 
\begin{equation}
\frac{\partial\mathbf{u}_{i}}{\partial\varphi}=\frac{1}{h_{i}}\frac{\partial h_{\varphi}}{\partial x_{i}}\mathbf{u}_{\varphi}-\delta_{i2}\sum_{k=1}^{3}\frac{1}{h_{k}}\frac{\partial h_{i}}{\partial x_{k}}\mathbf{u}_{k},\label{eq:lameDerivatives}
\end{equation}
with $i\in\left\{ 1,2,3\right\} $ and $h_{2}=h_{\varphi}$. For the
considered class of coordinate systems, none of the three metric scale
factors depends on the azimuthal variable and, by means of formula
(\ref{eq:lameDerivatives}), expression (\ref{eq:psiPrime}) reduces
to:
\begin{equation}
\boldsymbol{\psi}'\left(x_{1}',\varphi',x_{3}'\right)\sim\left[1-\varepsilon\frac{\partial}{\partial\varphi}+\varepsilon S_{z}\right]\left(\psi_{1}\mathbf{u}_{1}+\psi_{\varphi}\mathbf{u}_{\varphi}+\psi_{3}\mathbf{u}_{3}\right),\label{eq:psiPrimeSimplified}
\end{equation}
where $S_{z}$ represents a matrix operator which acts on the fundamental
column vectors: 
\begin{equation}
\mathbf{u}_{1}=\left[\begin{array}{c}
1\\
0\\
0
\end{array}\right];\quad\mathbf{u}_{\varphi}=\left[\begin{array}{c}
0\\
1\\
0
\end{array}\right];\quad\mathbf{u}_{3}=\left[\begin{array}{c}
0\\
0\\
1
\end{array}\right]
\end{equation}
and whose explicit form depends on the coordinate system:
\begin{equation}
S_{z}=\left[\begin{array}{ccc}
0 & -\frac{1}{h_{1}}\frac{\partial h_{\varphi}}{\partial x_{1}} & 0\\
\frac{1}{h_{1}}\frac{\partial h_{\varphi}}{\partial x_{1}} & 0 & \frac{1}{h_{3}}\frac{\partial h_{\varphi}}{\partial x_{3}}\\
0 & -\frac{1}{h_{3}}\frac{\partial h_{\varphi}}{\partial x_{3}} & 0
\end{array}\right].\label{eq:spinOperator}
\end{equation}
By direct calculation, we find:
\begin{equation}
B^{-1}\left[\begin{array}{ccc}
0 & -1 & 0\\
1 & 0 & 0\\
0 & 0 & 0
\end{array}\right]B=S_{z},
\end{equation}
where $B$ represents the basis change matrix from the curvilinear
coordinate system $\left\{ x_{1},\varphi,x_{3}\right\} $ to the Cartesian
one, $\left\{ x,y,z\right\} $. Therefore, $S_{z}$ exactly corresponds
to the third SO(3) generator in three-dimensional matrix representation. 

Since any arbitrary rotation of the vector function $\boldsymbol{\psi}$
can be expressed through the rotation operator, the following relation
holds:
\begin{equation}
\boldsymbol{\psi}'\left(x_{1}',\varphi',x_{3}'\right)=\exp\left(\varepsilon\hat{J}_{z}\right)\boldsymbol{\psi}\left(x_{1},\varphi,x_{3}\right)\sim\left[1+\varepsilon\hat{J}_{z}\right]\left(\psi_{1}\mathbf{u}_{1}+\psi_{\varphi}\mathbf{u}_{\varphi}+\psi_{3}\mathbf{u}_{3}\right).\label{eq:psiPrimeRot}
\end{equation}
If we compare equations (\ref{eq:psiPrimeSimplified}) and (\ref{eq:psiPrimeRot}),
we immediately obtain:
\begin{equation}
i\hat{J}_{z}=-i\frac{\partial}{\partial\varphi}+iS_{z},
\end{equation}
which is the sought expression for the third component of the vector
angular momentum operator (in $\hbar$ units). 

Let now $\boldsymbol{\psi}=\boldsymbol{\psi}_{m}$ be one of the three
vector functions $\left\{ \mathbf{M}_{m},\mathbf{N}_{m},\mathbf{L}_{m}\right\} $
defined in (\ref{eq:vectorFunctions}), with $g=g_{m}=\varPhi_{m}\left(\varphi\right)u\left(x_{1},x_{3}\right)$
representing a scalar function derived according to Section \ref{sec:Scalar}.
Owing to the fact that the three metric scale factors do not depend
on the azimuthal variable in the five considered systems, the partial
derivative $\partial/\partial\varphi$ commutes with each component
of the differential operators appearing in (\ref{eq:vectorFunctions}),
as can be shown explicitly by resorting to the formulas \cite{Morse1953}:
\begin{equation}
\nabla g=\sum_{i=1}^{3}\frac{\mathbf{u}_{i}}{h_{i}}\frac{\partial g}{\partial x_{i}};\quad\nabla\times\mathbf{C}g=\frac{1}{h_{1}h_{2}h_{3}}\left|\begin{array}{ccc}
h_{1}\mathbf{u}_{1} & h_{2}\mathbf{u}_{2} & h_{3}\mathbf{u}_{3}\\
\partial/\partial x_{1} & \partial/\partial x_{2} & \partial/\partial x_{3}\\
h_{1}C_{1}g & h_{2}C_{2}g & h_{3}C_{3}g
\end{array}\right|.
\end{equation}
Making use of this property and of equation (\ref{eq:infinitesimalTransf}),
it is easy to prove that: 
\begin{equation}
\boldsymbol{\psi}_{m}'\left(x_{1}',\varphi',x_{3}'\right)\sim\boldsymbol{\psi}_{m}\left(x_{1},\varphi,x_{3}\right)-im\varepsilon\boldsymbol{\psi}_{m}\left(x_{1},\varphi,x_{3}\right).
\end{equation}
Then, from comparison with (\ref{eq:psiPrimeRot}):
\begin{equation}
i\hat{J}_{z}\boldsymbol{\psi}_{m}\left(x_{1},\varphi,x_{3}\right)=m\boldsymbol{\psi}_{m}\left(x_{1},\varphi,x_{3}\right).
\end{equation}
\newpage


\providecommand{\newblock}{}

\end{document}